\documentclass[final,authoryear,3p,times]{elsarticle}

\usepackage[labelfont=bf]{caption}
\usepackage{graphics}
\journal{New Astronomy}

\begin{document}

\begin{frontmatter}


\title{Filament eruption in association with rotational motion near the filament footpoints}
\author[iia]{Sajal Kumar Dhara\corref{cor1}}
\ead{sajal@iiap.res.in}
\cortext[cor1]{Corresponding author Address: Indian Institute of Astrophysics, Bangalore 560034, India.
Tel.: +91 80 22531394; fax: +91 80 25534043}
\author[iia]{Ravindra B}
\author[iia,got]{and Ravinder Kumar Banyal}
\address[iia]{Indian Institute of Astrophysics, Bangalore 560034, India}
\address[got]{Institut f\"{u}r Astrophysik, Georg-August-Universitat G\"{o}ttingen, Friedrich-Hund-Platz
1, 37077 G\"{o}ttingen, Germany}

\begin{abstract}
The active region magnetic field surrounding the filament plays an important role in
filament formation, their evolution and disruption. We investigated a filament
eruption that occurred in southern hemisphere of the Sun on July 08, 2011 using AIA and
HMI data. The filament was located in a region close to the active region NOAA 11247
with its West-most footpoint anchored in the negative polarity plage region and
the East-most in the positive polarity
plage region. During observations, the magnetic flux was emerging in the active region and
also in the plage regions. The flux emergence was stopped in West-most footpoint of the
plage region about an hour before the filament eruption. A
converging motion was also observed for many hours in the Western footpoint of the filament.
The filament had left-handed twist and the net injected magnetic helicity
was positive in both footpoints. Both sign of magnetic helicity were observed in the
Western footpoint of the filament where the eruption has initiated.
Further, an anti-clockwise rotational motion was observed in both the footpoints
just after the onset of filament eruption which lasted for 6 min during the
eruption process.
The emerging flux, converging motion and injection of opposite magnetic helicity
could be responsible for destabilizing of the Western footpoint of the filament leading to eruption. The torque 
imbalance between the expanded portion
of the flux tube and the photosphere may have caused the rotation in the footpoint region
which changed the trend in the injected magnetic helicity after the filament eruption.
\end{abstract}
\end{frontmatter}

\section{Introduction}
 \label{S-Introduction}

Solar filaments are elongated thread-like, dark chromospheric structures with 
low temperature and  high density, suspended along the polarity inversion
line (PIL) between the regions of opposite magnetic polarities
\citep{Martin98a}. Filaments are formed in the active regions as well as in quiet magnetic
field regions and may last for several days. Most of them eventually disappear
with eruptions associated with flares and coronal mass ejections (CME)
\citep{Schmieder02, Gopalswamy03, Yan11}.

Twisted magnetic fields support the filaments in the corona.
The equilibrium loss initiated by kink-mode instability in twisted magnetic
fields is one of the leading mechanisms for filament destabilization and eruption
\citep{Sakurai76}. The kink-instability occurs when the twist of the
emerged flux ropes  exceeds a critical value, causing \emph{writhing} of
flux rope around its axis \citep{Fan05}. As a result, the flux rope can loose
its equilibrium and erupt \citep{Liu07}. By observing an active region
filament, \cite{Romano03} found that a total twist in one of the prominence threads
changed from 5-turns to 1-turn as it raised during filament activation.
They concluded that prominence was destabilized by kink-mode instability and  magnetic field later 
relaxed to a new equilibrium position.
\cite{Romano05} suggested that in one of the filament eruption events,
the injected magnetic helicity via the photospheric shearing motion exceeded the kink
instability threshold.

The filament eruption could be triggered by the magnetic
reconnection at low level in the solar atmosphere \citep{Contarino03, Contarino06}.
In the photosphere it is seen as a cancellation of magnetic features
\citep{Priest94}.  Many times it has been observed that magnetic
flux cancellation occurs at the photosphere near the PIL \citep{Martin98a}.
If the flux cancellation at the photosphere continues after the flux rope
has been formed, it may result in instability leading to eruption; as formulated
by \cite{vanBallegooijen89, Amari03, Amari11}. It has also been observed that flux cancellation
at the PIL lead to the formation of X-ray sigmoid which eventually triggers the
CME \citep{Green11}. A successive reconnection in the
coronal arcade can change the configuration such that filament below the arcade
can no longer be sustained. This destabilization can cause the eruption of the filament
\citep{Zuccarello07}. ``Tether-cutting'' mechanism is another example of
magnetic reconnection where strapping of magnetic tension force is released by internal
reconnection above the PIL \citep[e.g.,][]{Moore01} to
destabilize the filament. But, \cite{Aulanier10} found that magnetic flux cancellations at the
photosphere and the tether-cutting reconnection at the coronal heights do not initiate
the CMEs in bipolar magnetic field configurations.  However, they are essential to buildup 
flux ropes in the pre-eruptive stage. Subsequently, the flux
rope rises to a height at which the torus instability can set in to cause the eruption.

Sunspot rotation in the vicinity of filament can also be responsible for the formation and
ejection of active region filament \citep{Yan12}. \cite{Zuccarello12b} have observed a
B7.4 class flare associated with filament eruption which occurred in the active region. By examining
the magnetic field configuration and photospheric velocity maps they concluded
that a shearing motion of the magnetic field lines could increase the axial
field of the filament, thus bringing the flux rope to a height where the torus
instability criteria is met to favor the eruption.

In this paper, we present the study of a filament eruption that occurred in the vicinity of
the active region NOAA 11247 that was observed in different wavelength regimes. The filament
eruption which occurred at about 23:20~UT on July 08, 2011 was followed by B4.7 class flare
starting at $\sim$ 00:45~UT.
Prior to eruption, a flux emergence in the vicinity of filament footpoint and 
converging flow were observed in 
two footpoints of the filament at the photosphere. These footpoints of
the filament were rooted in the West-most plage region. Just after the filament activation, 
a rotational motion was also observed at the footpoint locations. In the next Section, we
describe the data preparation and the analysis method used. The observational
results starting with flux emergence, convergence flow around
the filament footpoint, filament eruption and subsequent rotation in the footpoint
are presented in Section \ref{section3}. Finally, in the
last Section, we discuss the importance of the flux emergence and converging motion
associated with filament destabilization and eruption. We also give plausible explanation for the
observed rotation in the footpoints.

\section{Observation and method}
     \label{section2}

Atmospheric Imaging Assembly \citep[AIA;][]{Lemen12} on board the Solar Dynamics Observatory
(SDO) provides images of the Sun
in ultra violet (UV) and extreme ultra violet (EUV) wavelengths that cover the regions
from the photosphere to the coronal level. The pixel resolution of AIA is 0.6$^{\prime\prime}$
while the field-of-view is about 1.4R$_\odot$. We used EUV data from AIA at
171, 193 and 304~\AA~wavelengths (image cadence of 12~sec) to study the filament eruption in detail.
The acquired data set starting from 15:00 UT (July 08, 2011) to 04:00 UT (July 09, 2011)
covers the entire filament eruption event. The level-1 data was upgraded to the level 1.5
using the software `AIA$\_$PREP.PRO' available in {\emph solarsoft} routine. The region of interest
was tracked using the `DEROT$\_$MAP.PRO'. The tracked region data cube was prepared and used to
study the dynamics of the filaments at the coronal level.

Complementary to the coronal data sets, we also obtained a few full-disk H$_{\alpha}$ images
from the Big Bear Solar Observatory
to study the morphology of the filament at the chromospheric heights.
These H$_{\alpha}$ images were acquired at BBSO \citep{Denker99} with 2k$\times$2k pixel
CCD camera having a pixel resolution of 1$^{\prime\prime}$. Similar to AIA data, we also
extracted the region of interest in H$_{\alpha}$ images by tracking it in heliographic
co-ordinate system. However, due to their limited availability,
we used them only for the morphological study of the filament in the chromosphere.

Helioseismic and Magnetic Imager \citep[HMI;][]{Schou12} provides the line-of-sight magnetogram at
a cadence of 45~sec with a spatial resolution of 0.5$^{\prime\prime}$ per pixel. We have
obtained the line-of-sight magnetograms for about 2 days starting from Jul 08, 2011. The
data has been interpolated to the AIA pixel resolution. Later, the obtained data
has been tracked over the region of interest, corrected for the line-of-sight effect by
multiplying 1/cos$\theta$, where $\theta$ is the heliocentric angle. We averaged 4 magnetograms
to reduce the noise level to 10~G. These magnetograms were then used to study the evolution of
magnetic fields in and around the filament at the photospheric level.

Apart from these data sets, we also acquired the continuum intensity images
from the HMI telescope for which the cadence and pixel resolution are same as the
line-of-sight magnetograms. The obtained images have been tracked over the region of
interest as has been done for the other data set mentioned above. The
`rotation-corrected' images were then used to determine the velocity of small features near
the filament footpoints.

\renewcommand{\figurename}{Fig.}
\begin{figure}
\begin{center}
\includegraphics[width=0.515\textwidth,clip=]{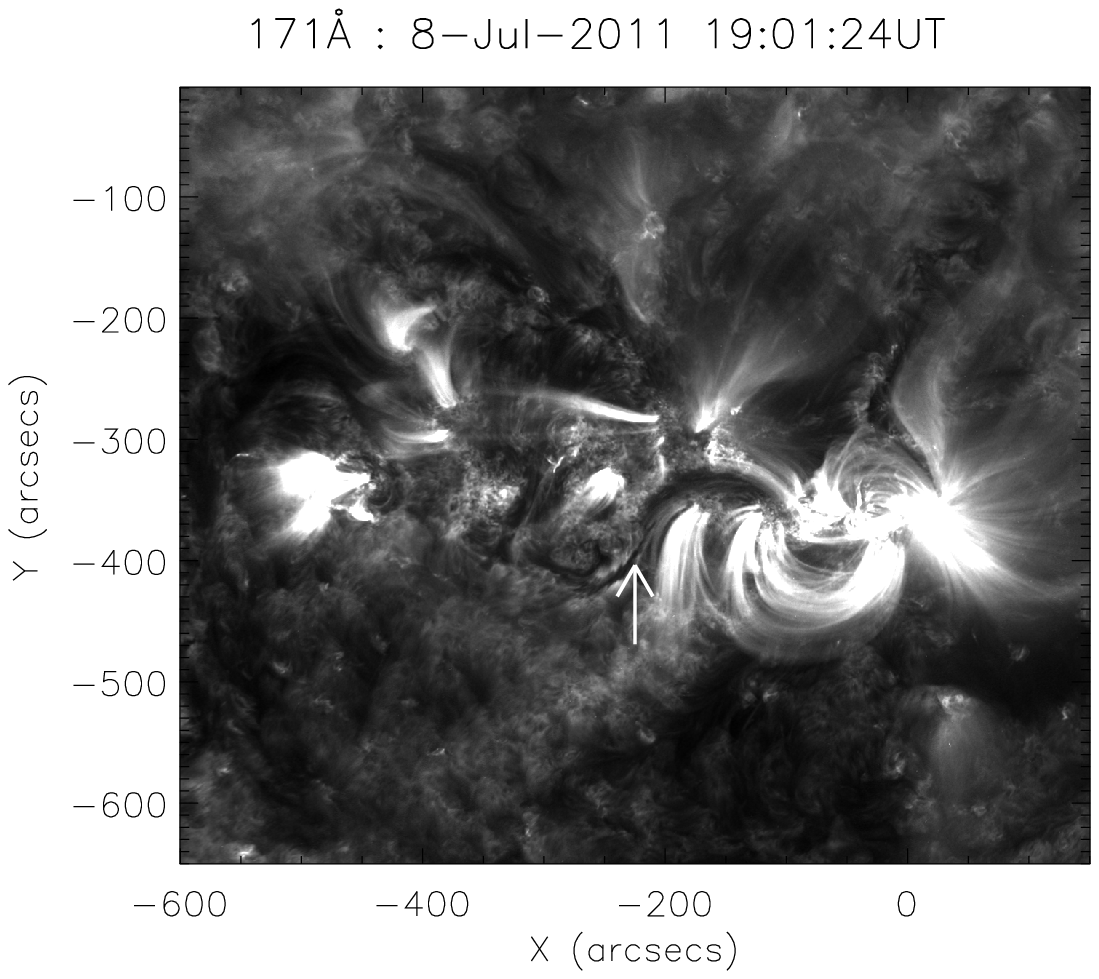}\hspace{-0.3in}\includegraphics[width=0.515\textwidth,clip=]{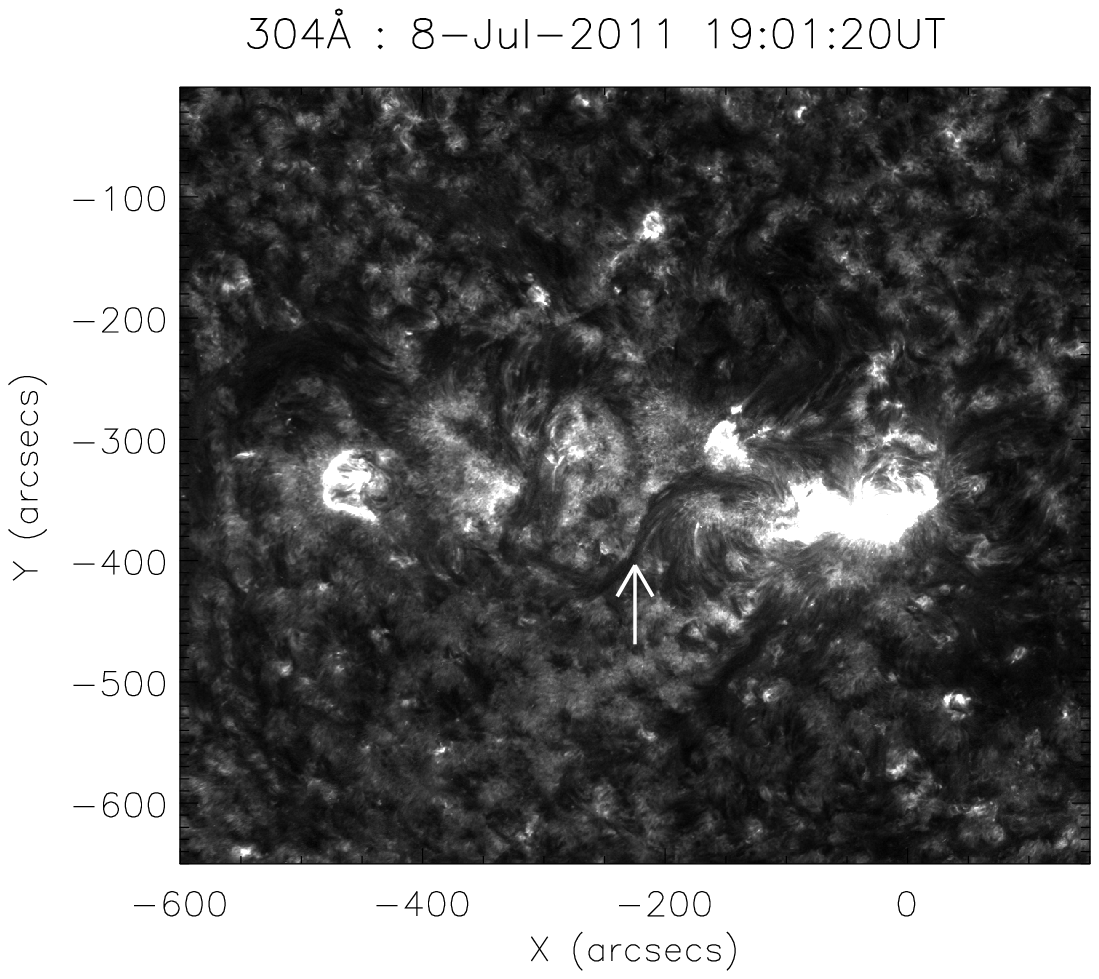} \\
\vspace{-0.2in}\includegraphics[width=0.515\textwidth,clip=]{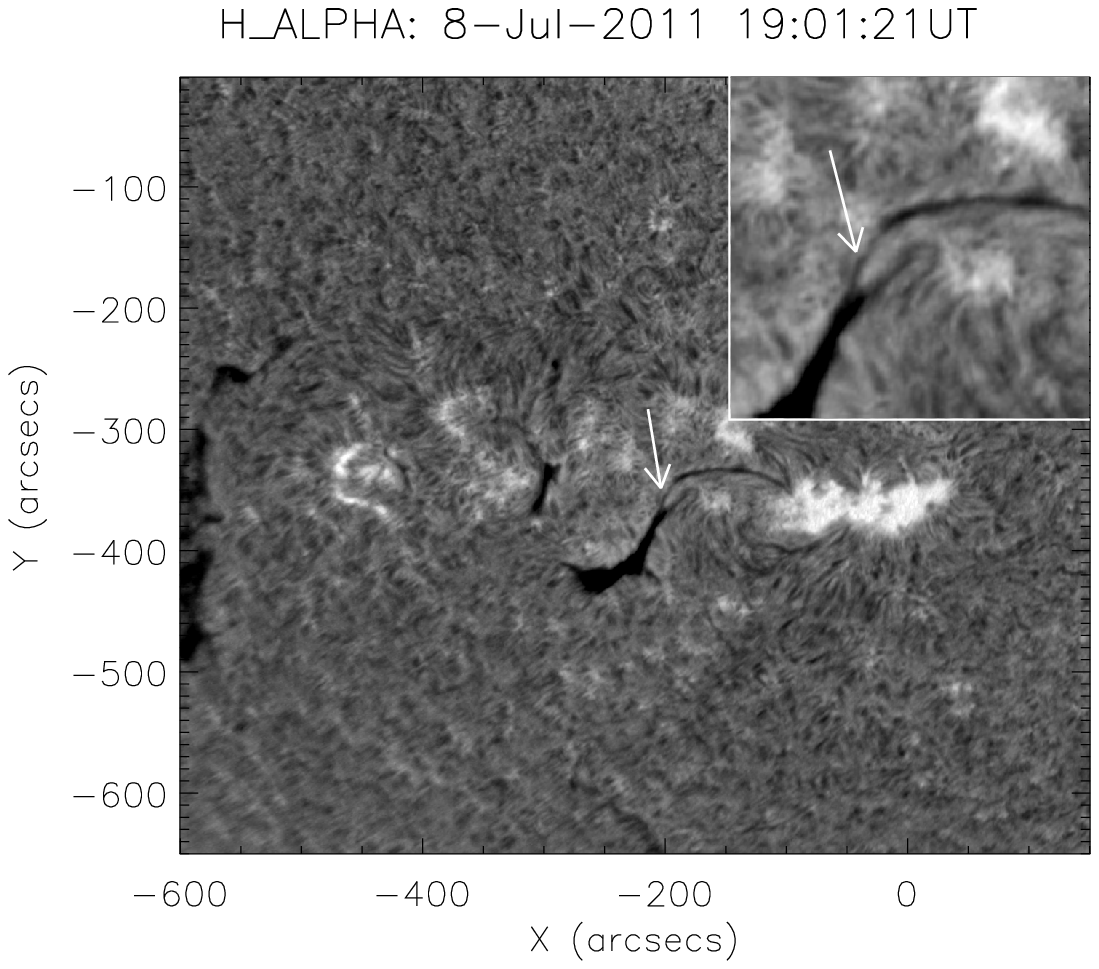}\hspace{-0.3in}\includegraphics[width=0.515\textwidth,clip=]{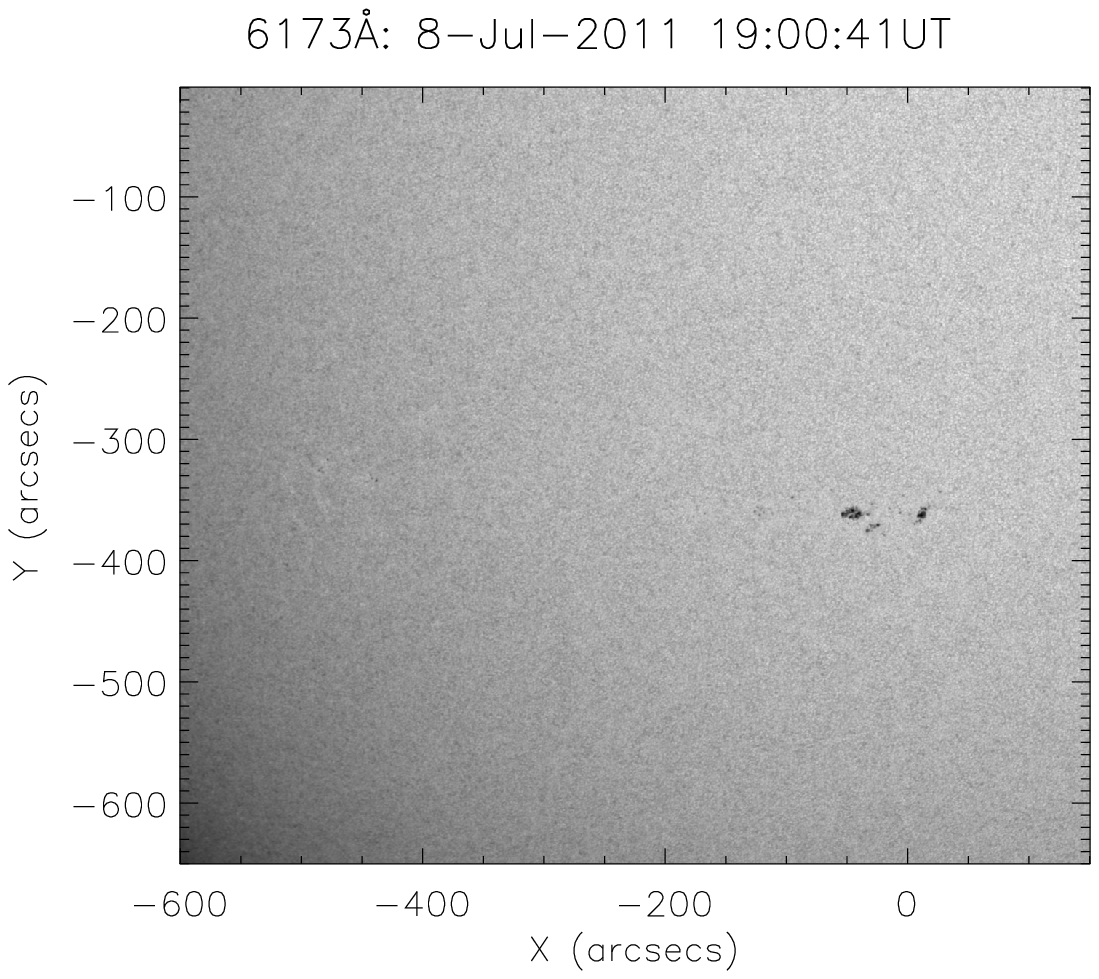} \\
\end{center}
\caption{Observed filament in the 171~\AA~channel of the coronal image
(top left), 304~\AA~of the transition region image (top right), H$_{\alpha}$
chromospheric image (bottom left) and photospheric continuum image (bottom right).
The field-of-view is same in all the four images. In top two
images, the filament location is indicated by white arrow. The inset image on the top right 
of H$_{\alpha}$ image shows 
the magnified portion of the bifurcated thin thread like
structure of the filament.}
\label{fig:1}
\end{figure}

\section{Results}
\label{section3}

\subsection{Filament observations at various heights}
  \label{fcorona}

The filament was located just outside the active region NOAA 11247  at a latitude of
19$^{\circ}$ in the southern hemisphere. The filament was visible on the Eastern limb when the active
region turned towards Earth on July 05, 2011. During the filament eruption the active region was
located at a longitude of 14$^{\circ}$ East of the central meridian.
Figure~\ref{fig:1} shows the active region as observed in different wavelengths from
the corona to the photosphere. The `S-shaped' filament is vividly seen in coronal 171~\AA~image
(top-left) and in the higher chromosphere-transition region image taken in 304~\AA~wavelength
(top-right). The filament in EUV wavelength is observed as dark feature.
In the H$_{\alpha}$ image (bottom-left) the filament is visible as a thick dark
structures located away from the active region and thin thread like structures extended up to 
bright plage region. The thin thread like structure bifurcates into two parts. The one
part extends up to the plage region and the other one terminates somewhere in the quiet sun
region. The bifurcation can be clearly seen in the enlarged portion displayed in the same
H$_{\alpha}$ image, indicated by an arrow. Note that the thin thread like structure
seen in H$_{\alpha}$ image
appears as thick dark structure in the coronal image. This bifurcation of the filament in the
Western footpoint is not visible in 304~\AA~image rather a continuous thick dark structure which
ends in the Western plage region is seen.
In the H$_{\alpha}$ image the filament exhibits a discontinuity at a
location of -300 to -350 arcsec along the horizontal axis. The discontinuity in filament structure is also seen 
in coronal image (top-left) close to the East-side that is indicated by white arrow.  The photospheric image
(bottom-right) shows two sunspots in the AR 11247 that are located in the vicinity of the
filament structure.

\begin{figure}
\begin{center}
\includegraphics[width=0.515\textwidth,clip=]{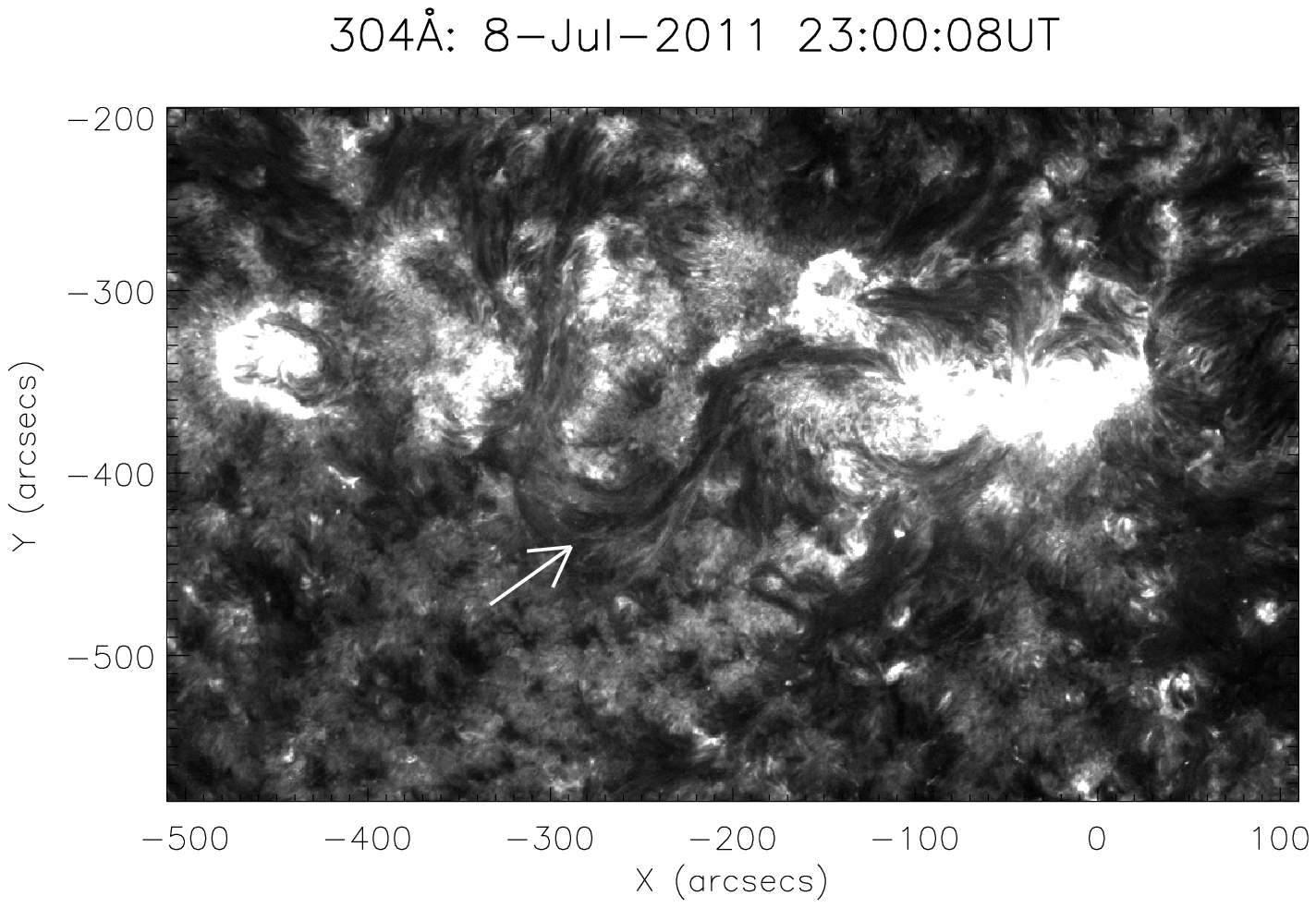}\hspace{-0.4in}\includegraphics[width=0.515\textwidth,clip=]{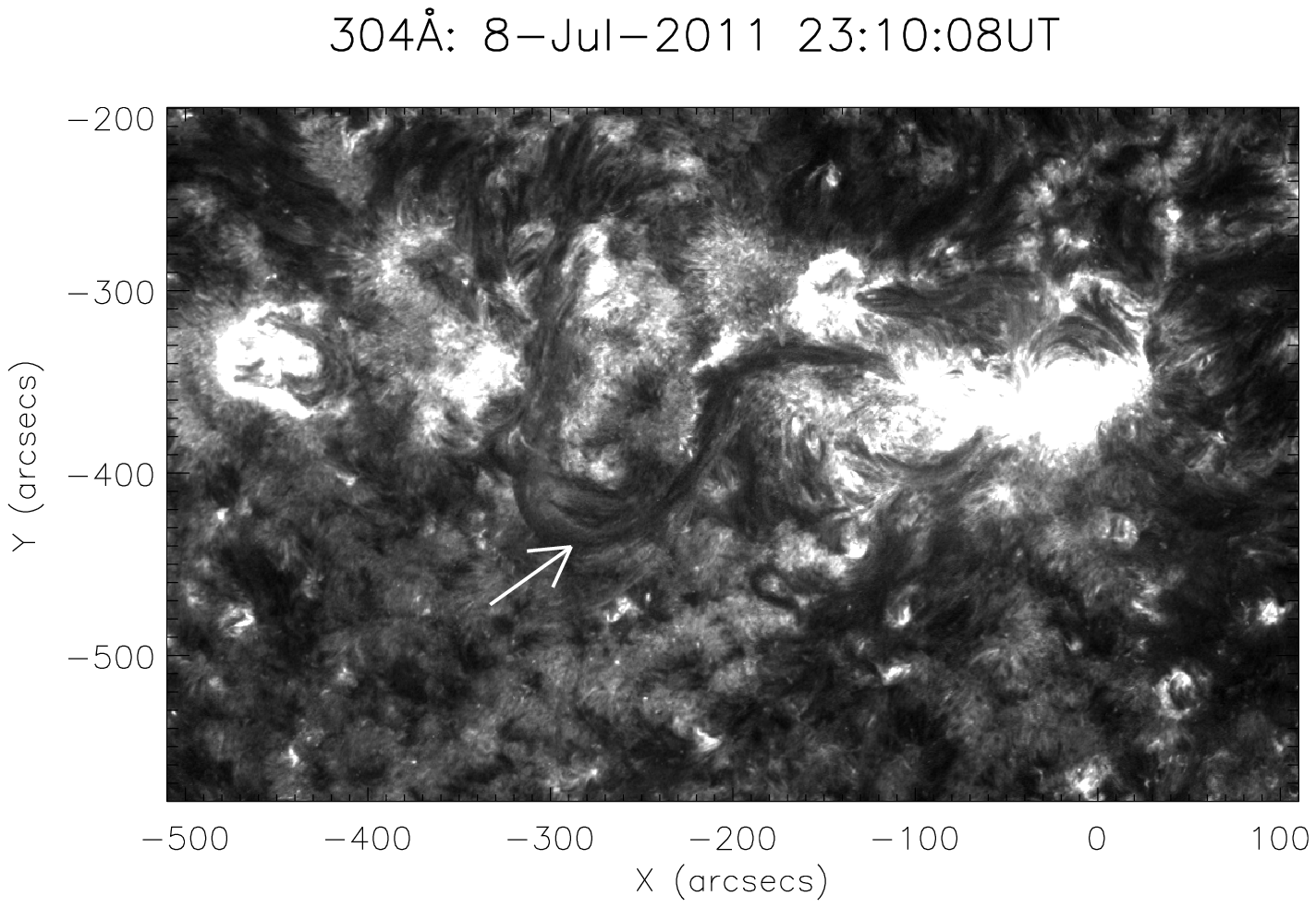} \\
\vspace{-0.3in}\includegraphics[width=0.515\textwidth,clip=]{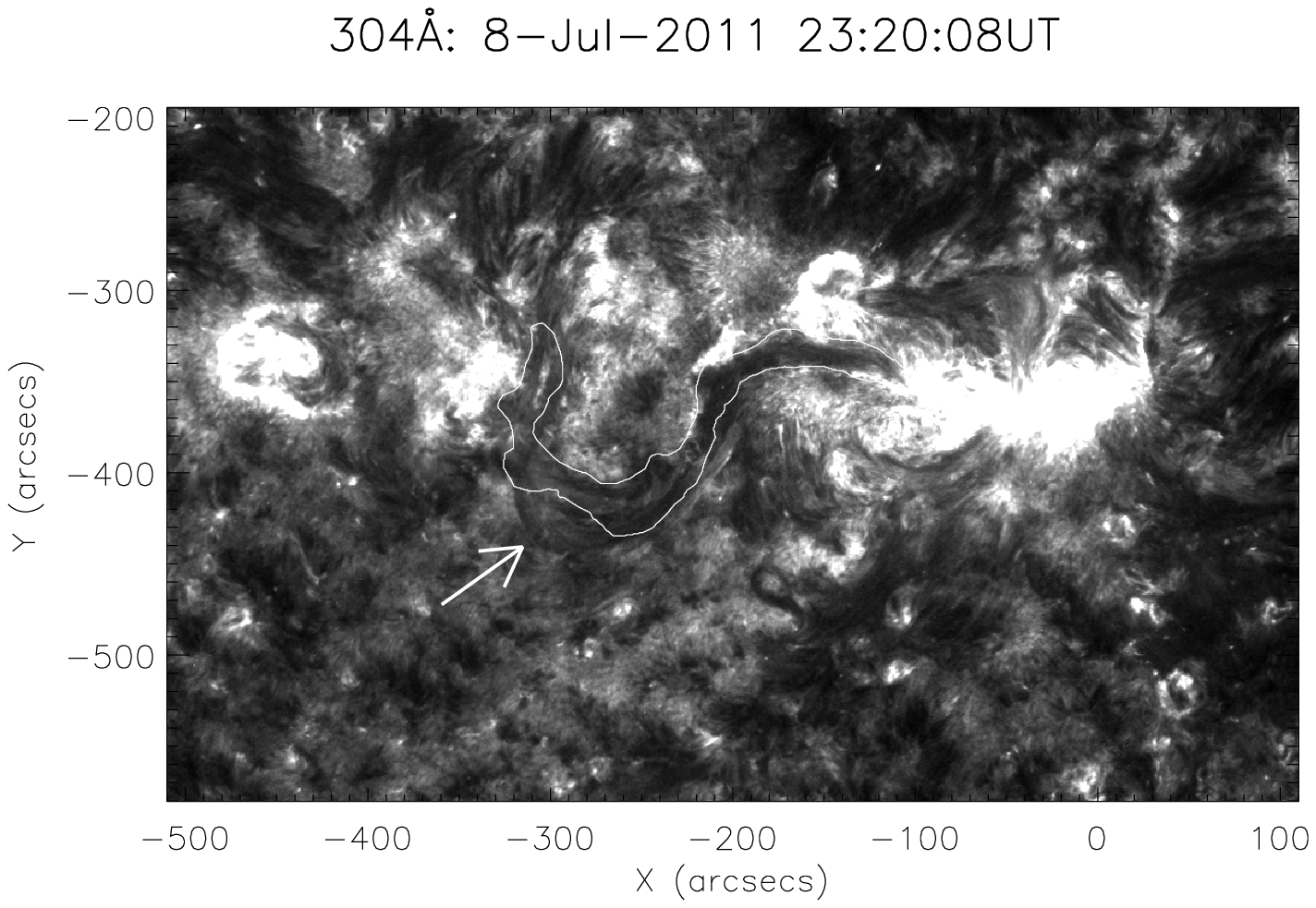}\hspace{-0.4in}\includegraphics[width=0.515\textwidth,clip=]{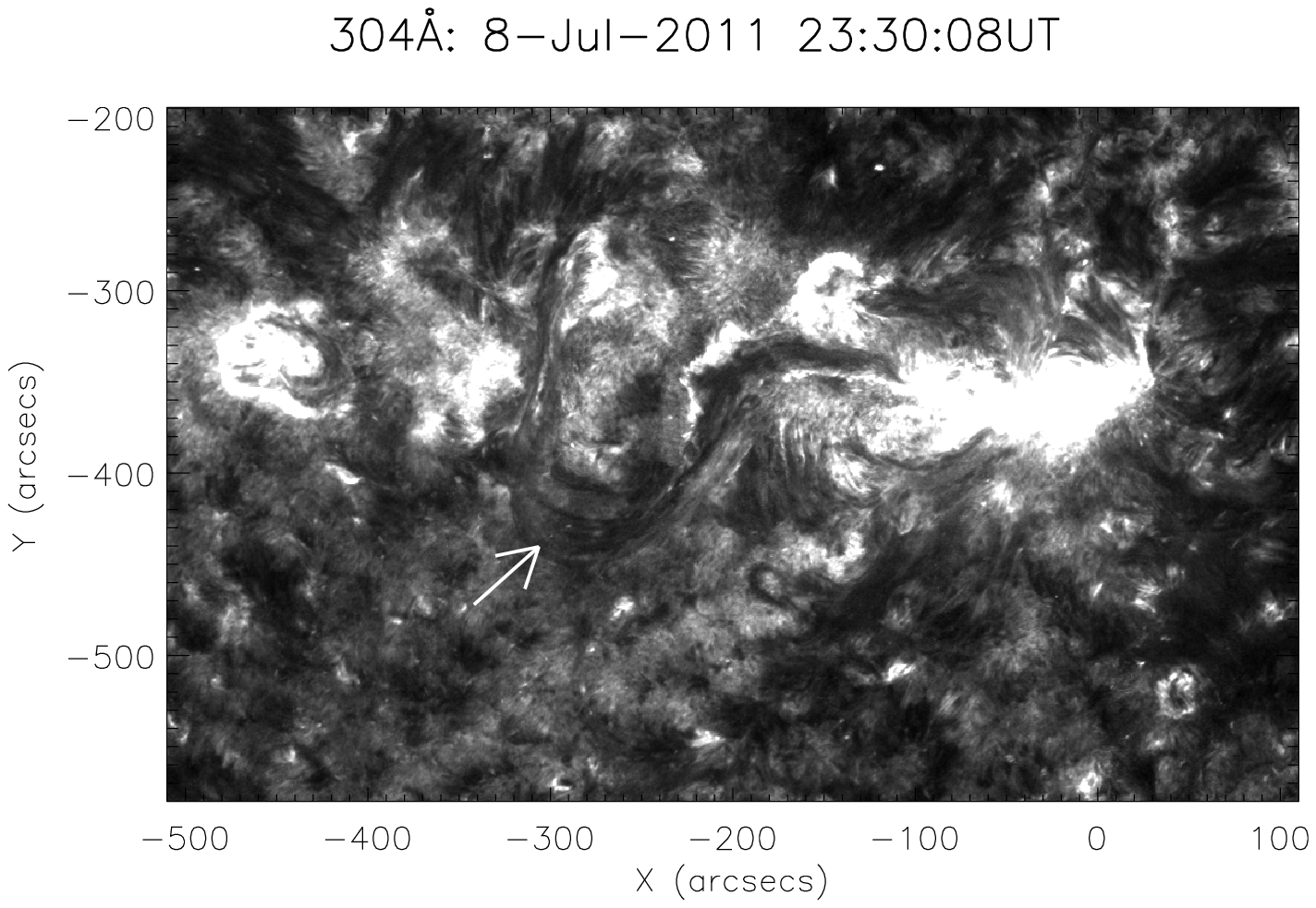} \\
\vspace{-0.3in}\includegraphics[width=0.515\textwidth,clip=]{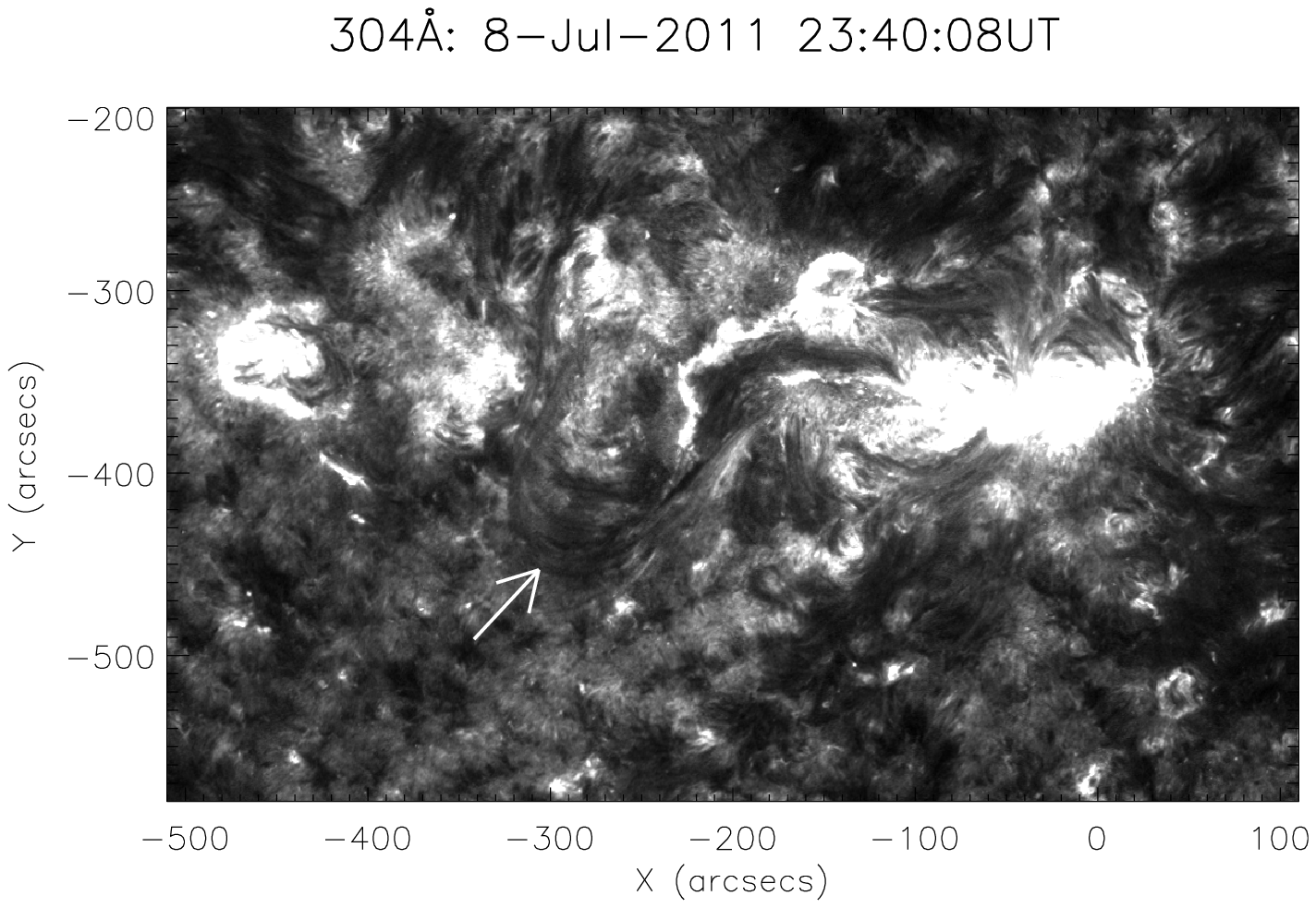}\hspace{-0.4in}\includegraphics[width=0.515\textwidth,clip=]{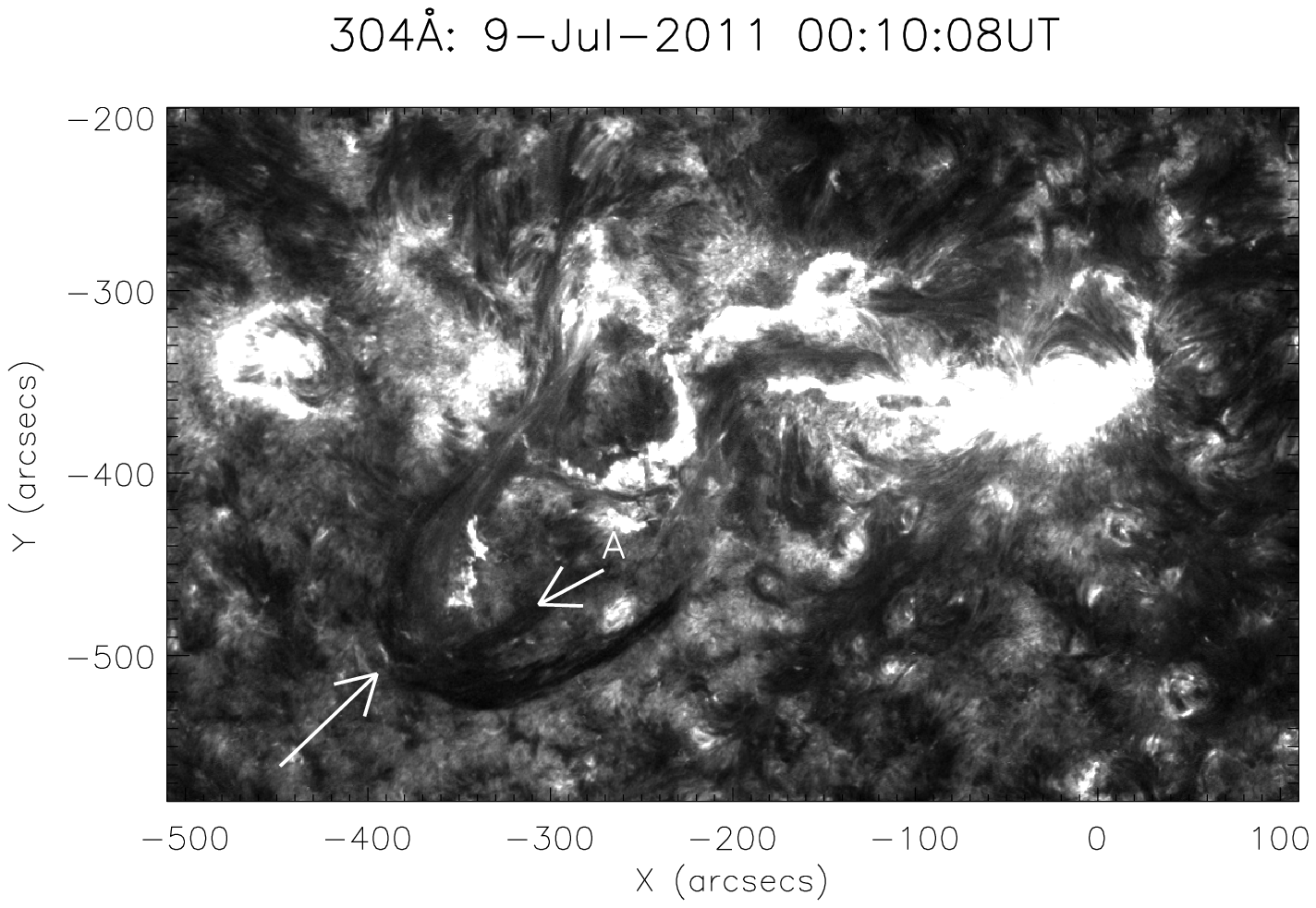} \\

\end{center}
\caption{A time sequence of erupting filament is shown at different epoch in
 He~II~304~\AA~images. The white arrow indicates the position
of the filament and the letter `A' (bottom-right) with an arrow shows the position of a
small portion attached to the bottom side of the filament that took
an anomalous path while erupting. In the middle-left image the contour of the filament
channel is overlaid on the erupting filament.}
\label{fig:2}
\end{figure}

\begin{figure}
 \begin{center}
\includegraphics[scale=0.8]{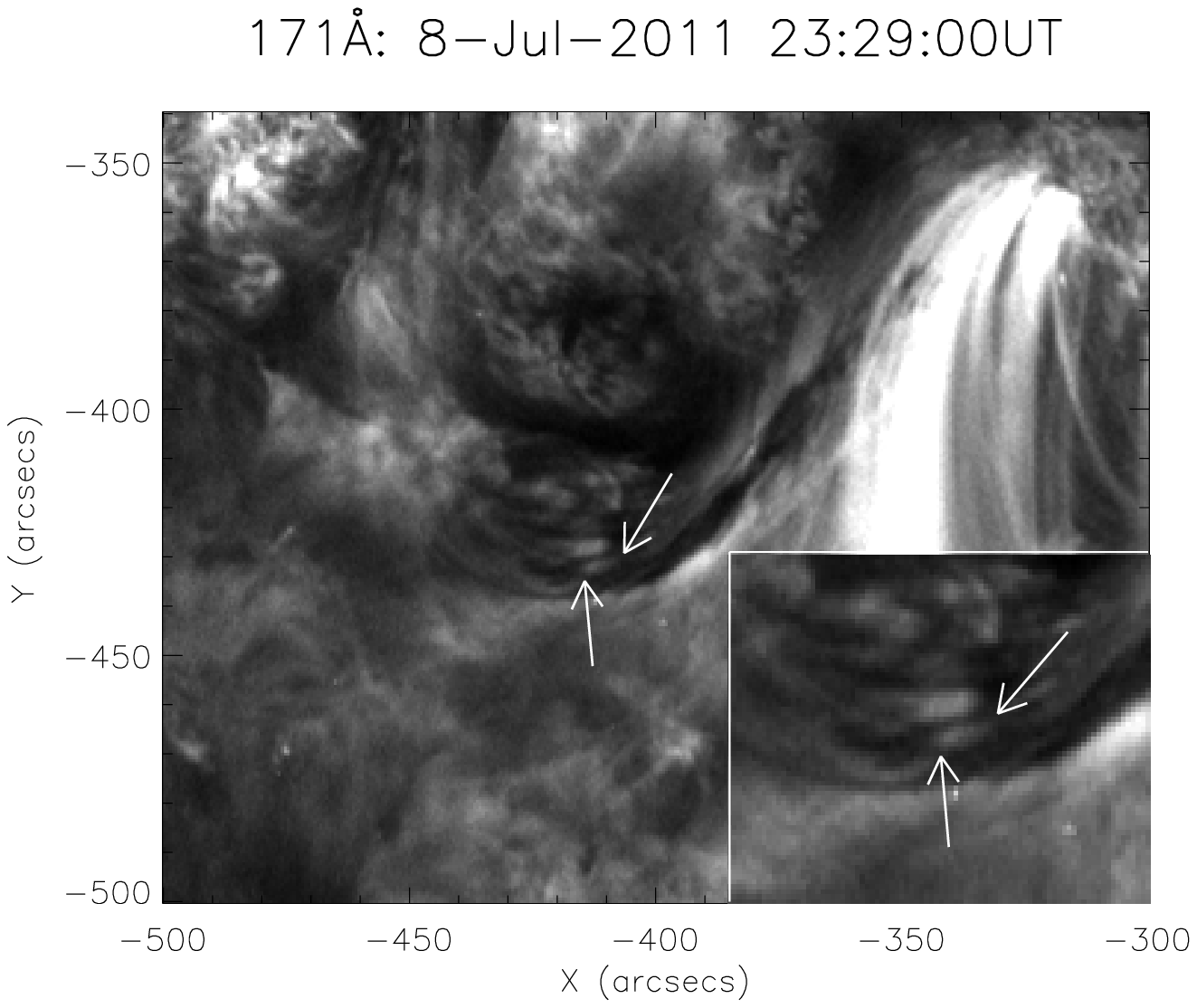} \\
\end{center}
\caption{Erupting filament shown in 171~\AA~image taken by AIA/SDO. The arrows indicate crossing of dark threads
over the bright threads. The box on the bottom right of the image shows the zoomed in version
of the cross over threads.}
\label{fig:3}
\end{figure}

\subsection{Filament eruption}
\cite{Rust96} suggested that the solar filaments and its neighboring structures exhibit
the same sign of twist. The filament studied here appeared as `S-shaped' structure. This
suggests that the filament has left-handed chirality with positive helicity sign \citep{Martin03}.
Also, the axial field of the filament is left-bearing for an observer looking
at the filament from North polarity side of magnetic field (c.f. Fig. \ref{fig:5})
and hence the filament is sinistral \citep{Martin98b}.

Figure~\ref{fig:2} shows a sequence of 304~\AA~images while the filament is in
erupting phase. In each of these images, the location of the filament is indicated
by an arrow. In the middle-left panel image of Fig.~\ref{fig:2} we overlaid the contour
of the filament extracted from an image obtained an hour before the filament eruption.  It is evident
that filament had started erupting at this stage. 
The region which is producing out 
of the contoured region
(shown by an arrow) suggests the initiation of the filament eruption.
The filament started
to erupt at $\sim$ 23:20~UT on July 08, 2011 and became elongated. Later, it 
completely disappeared from the field of view at around
00:29~UT on July 09. In middle portion of the filament a discontinuity was
observed (see Section \ref{fcorona}). During the eruption this portion 
was seen as a tail shown by a letter `A' with an arrow in the bottom-right
image of Fig.~\ref{fig:2}. The filament eruption was followed by 
two-ribbon flare (B4.7~class), observed at $\sim$ 00:45~UT.

The erupting filament (see Fig. \ref{fig:3}) shows a crossing of bright and dark
threads in 171~\AA~images.  The crossing of dark features over the bright
regions is easily identified in the zoomed image where a bright feature in the background is seen 
going from right to left while the dark feature in the foreground crosses it from left
to right. This crossing corresponds to the positive mutual (type I)
helicity \citep{Chae00}. Overall, the filament exhibits positive sign of helicity
when it is quiet and also shows the same sign while it is erupting.

\begin{figure}
 \begin{center}
\includegraphics[scale=0.6]{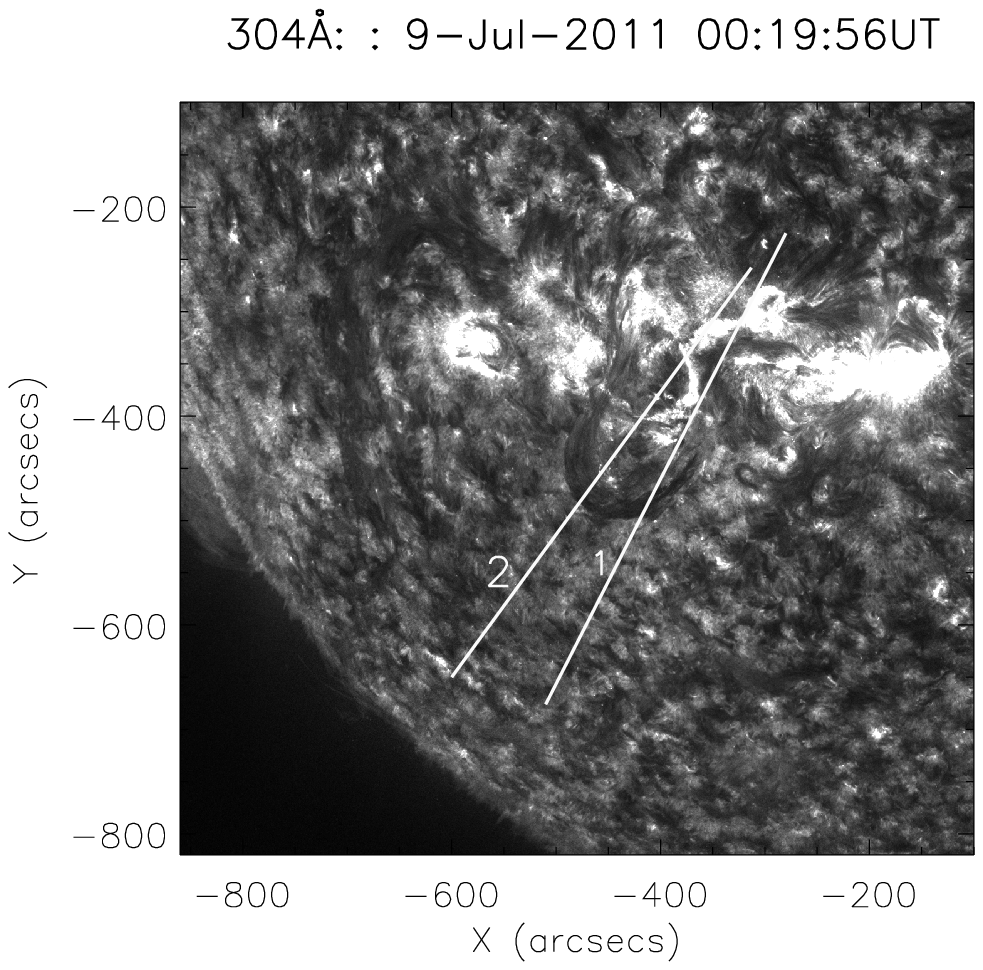} \\
\vspace{-0.2in}\includegraphics[width=0.515\textwidth,clip=]{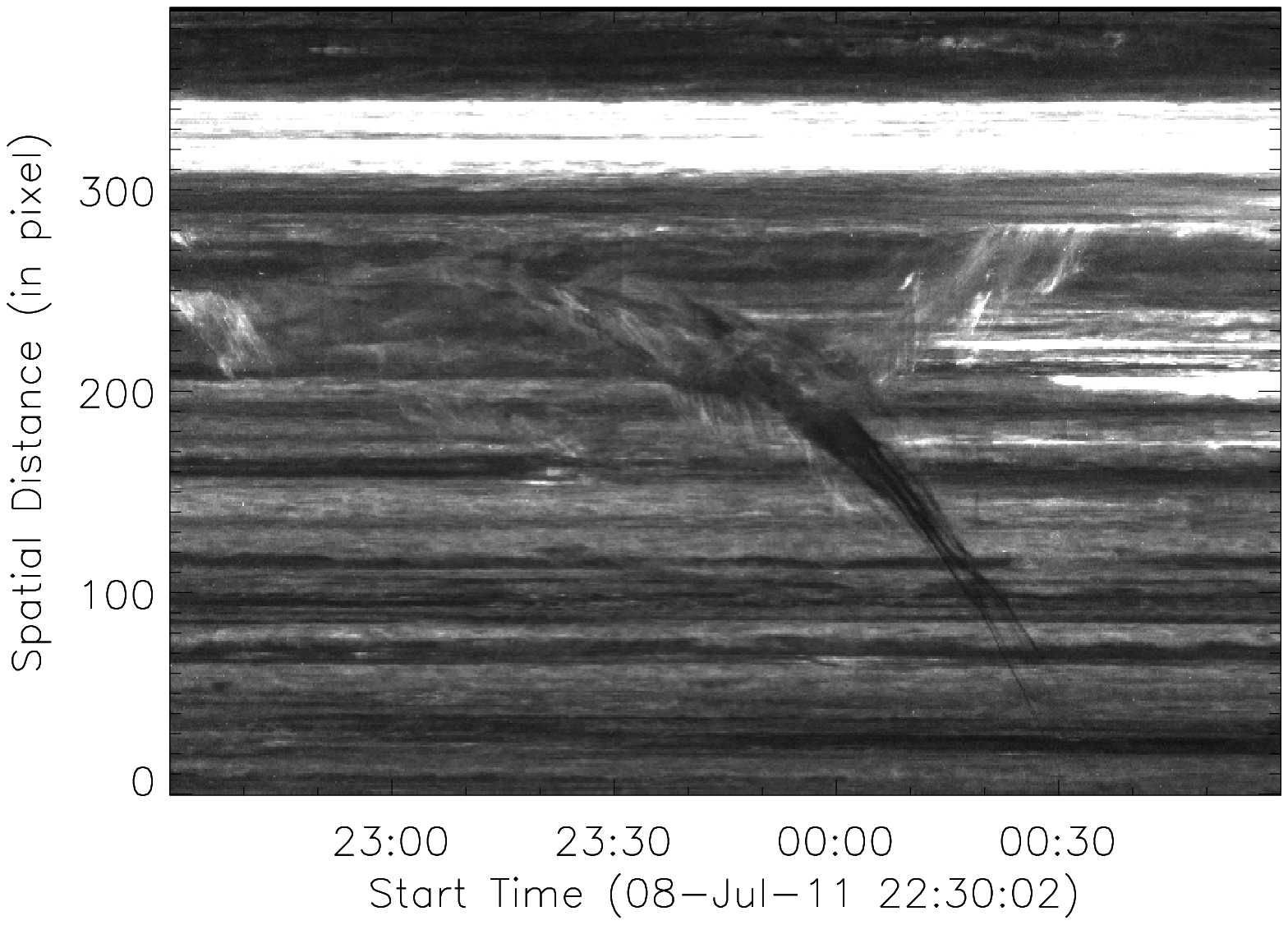}\hspace{-0.2in}\includegraphics[width=0.515\textwidth,clip=]{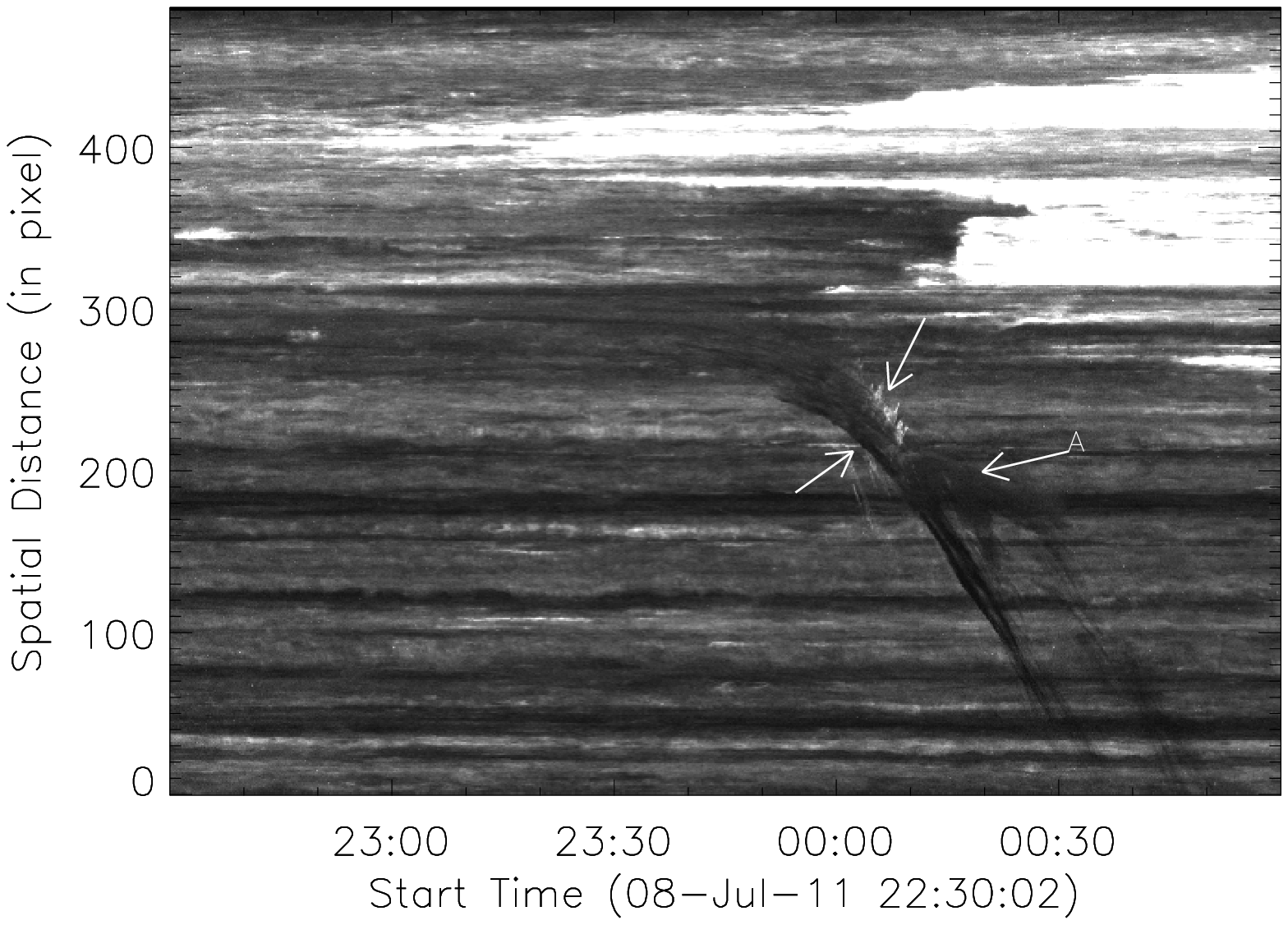} \\
\end{center}
\caption{Top: The filament is shown in 304~\AA~image. Two white slits overlaid
on the image represent the position from which the space-time maps have been generated. The slit number 1
crosses the Western footpoint of the filament. The slit number 2 crosses
the filament midway between the footpoints.
Bottom: The space-time map of filament eruption for slit number 1 (left) and for slit number
2 (right). A dark bended portion of the filament
represents the path of the filament while erupting. The white arrow followed by letter `A' (bottom-right image)
showing portion of the filament which took a anomalous path. Other two white arrows
shows the location of surge in the space-time map.}
\label{fig:4}
\end{figure}

The space-time map has been made using 304~\AA~images to study the evolution of
erupting filament. The top image in Fig.~\ref{fig:4} shows two slit positions from
which space-time maps were extracted in 304~\AA~wavelength. The slit position 1 provided
the space-time map for the Western footpoint of the filament (bottom-left). The bottom-right
image shows the space-time map for the erupting filament extracted from the slit position 2. 
The bottom-left space-time map shows that
the filament activation started at about 23:20~UT at the Western footpoint.
The space-time maps show the path of the erupting filament which is curved.

A brightening on the top and bottom part of the erupting filament from
23:55 to 00:12~UT (shown by two inward arrows) can be seen in the bottom-right space-time map. 
This type of brightening,  also observed simultaneously in other wavelengths (171~\AA~ \& 193~\AA~), is caused by a surge
which started from the footpoint of the filament and moved faster than the erupting filament.

A small portion of the filament marked as `A' in Fig.~\ref{fig:2} (bottom-right)
can also been seen in space-time map (shown by the letter `A' with an arrow).
Unlike normal curved path, this portion took a different
trajectory and moved away from the main filament channel.
This anomalous path is not visible in 171~\AA~images, suggesting that it may have collapsed
before entering the corona. This portion arose from discontinuity in the filament region (see Section \ref{fcorona}).
The projected velocity of the erupting filament is 56~$\pm$~1.4~km~s$^{-1}$. This is obtained by extracting the data points on the curved
portion of the path of the erupting filament in the space-time map and then fit a linear
least square fit to the data points. The filament was not visible in coronal images after 00:30~UT on July 09, 2011.

\begin{figure}
\begin{center}
\includegraphics[width=0.8\textwidth,clip=]{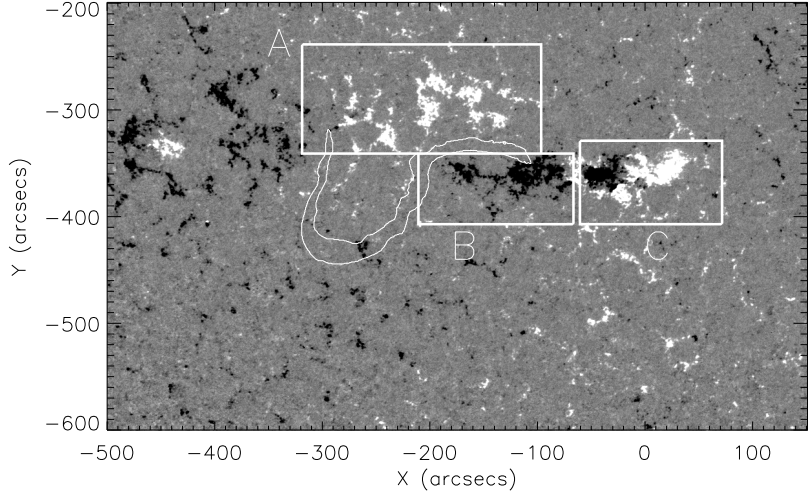}
\end{center}
\caption{The line-of-sight magnetogram showing the location of the active region and plages.
A white contours overlaid upon the magnetogram indicates the filament boundaries.
The box A and B represent the plage region and box C represents the emerging flux region.}
\label{fig:5}
\end{figure}

\begin{figure}
\begin{center}
\includegraphics[width=0.5\textwidth,clip=]{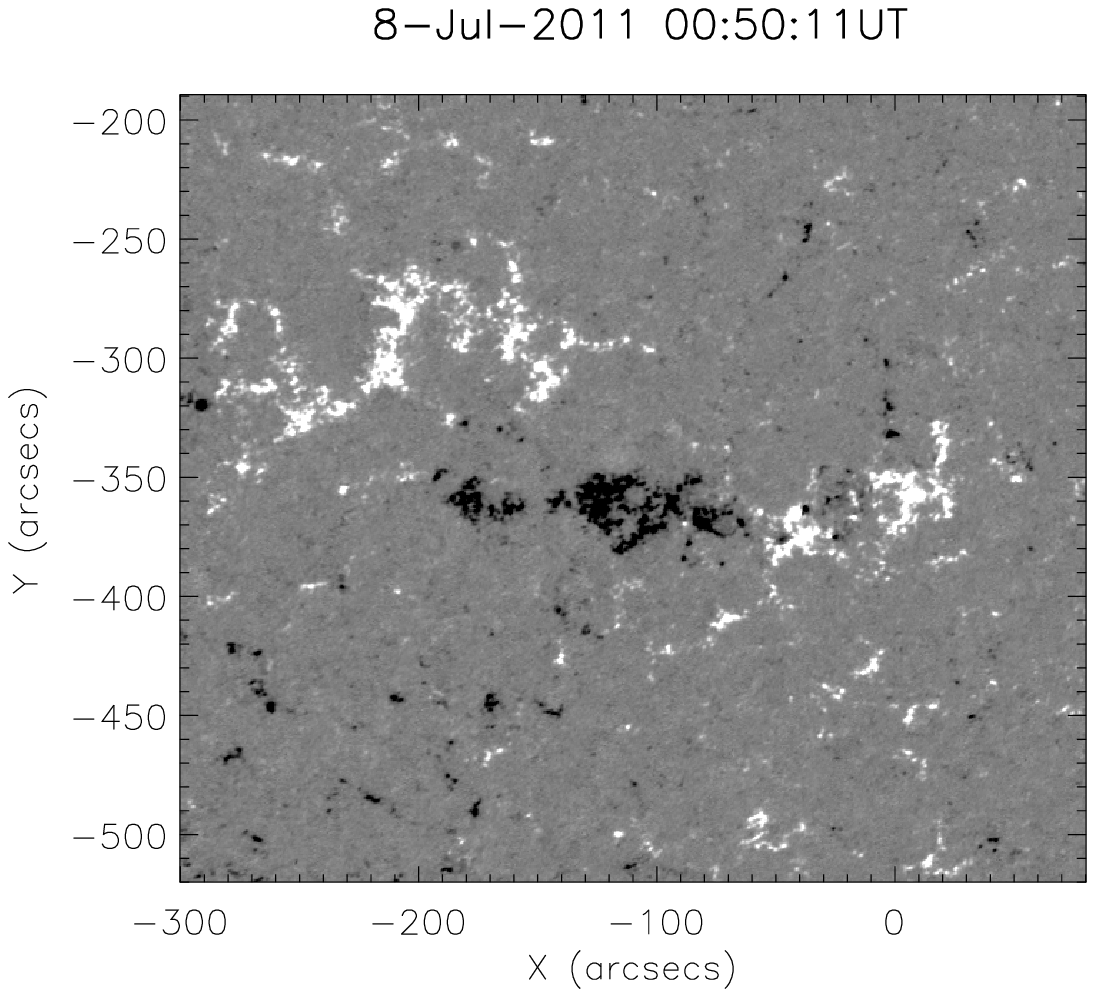}\hspace{-0.3in}\includegraphics[width=0.5\textwidth,clip=]{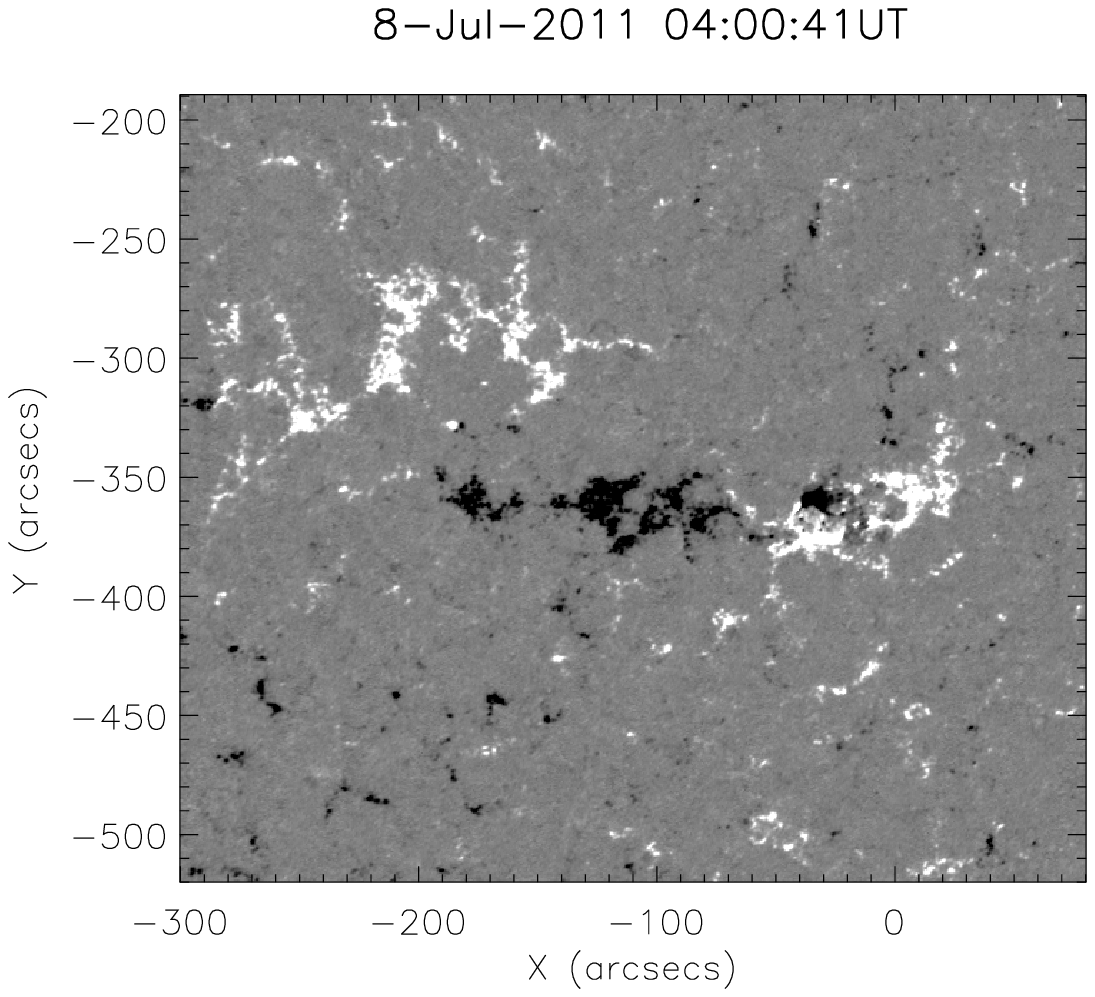} \\
\vspace{-0.2in}\includegraphics[width=0.5\textwidth,clip=]{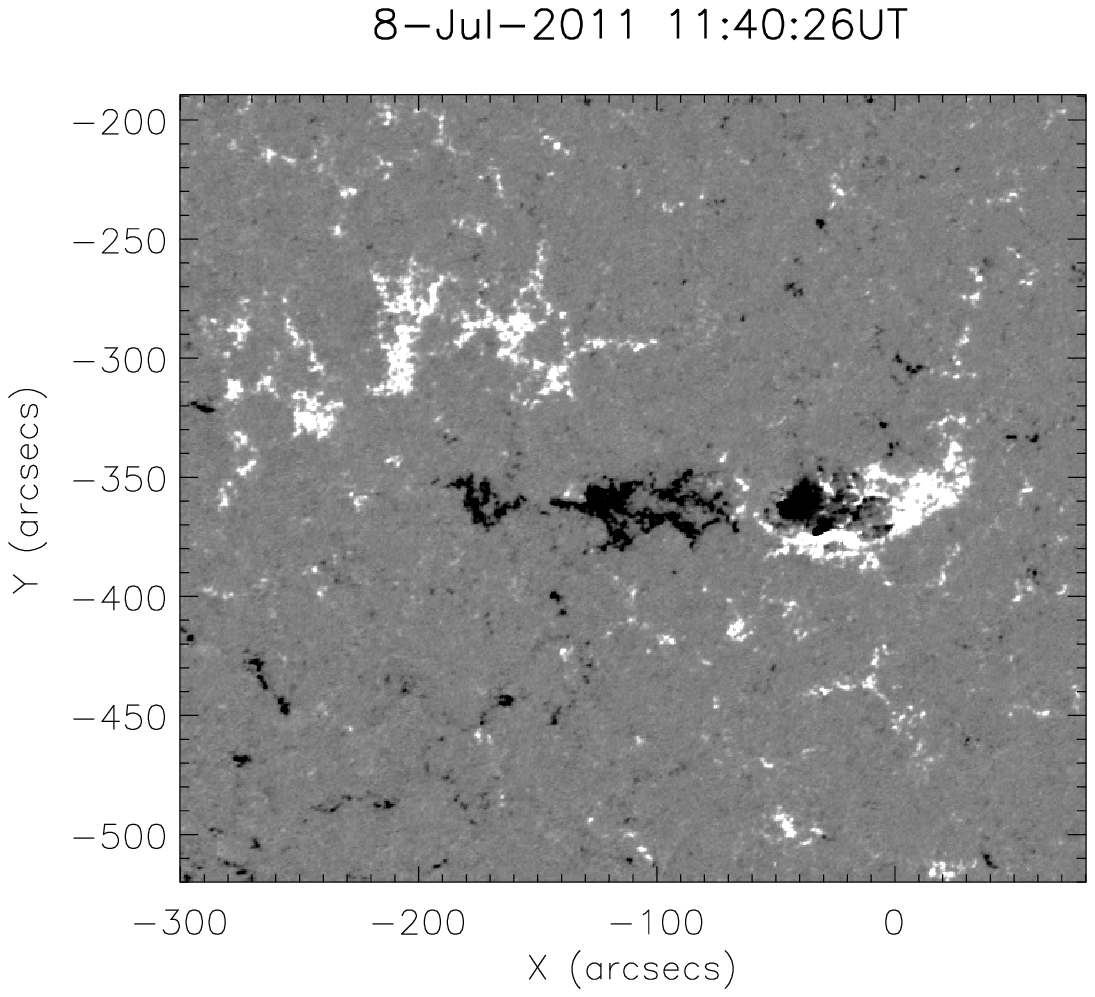}\hspace{-0.3in}\includegraphics[width=0.5\textwidth,clip=]{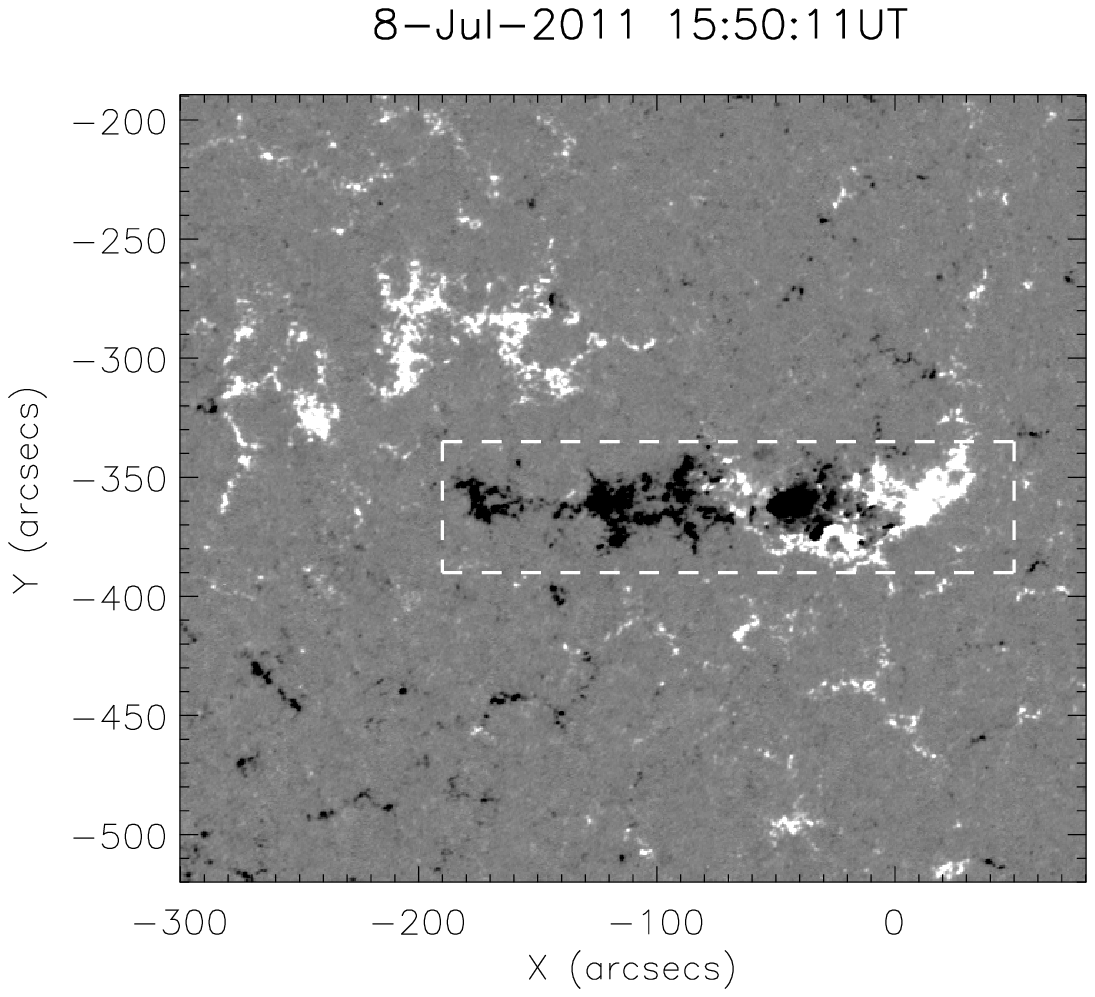} \\
\vspace{-0.2in}\includegraphics[width=0.5\textwidth,clip=]{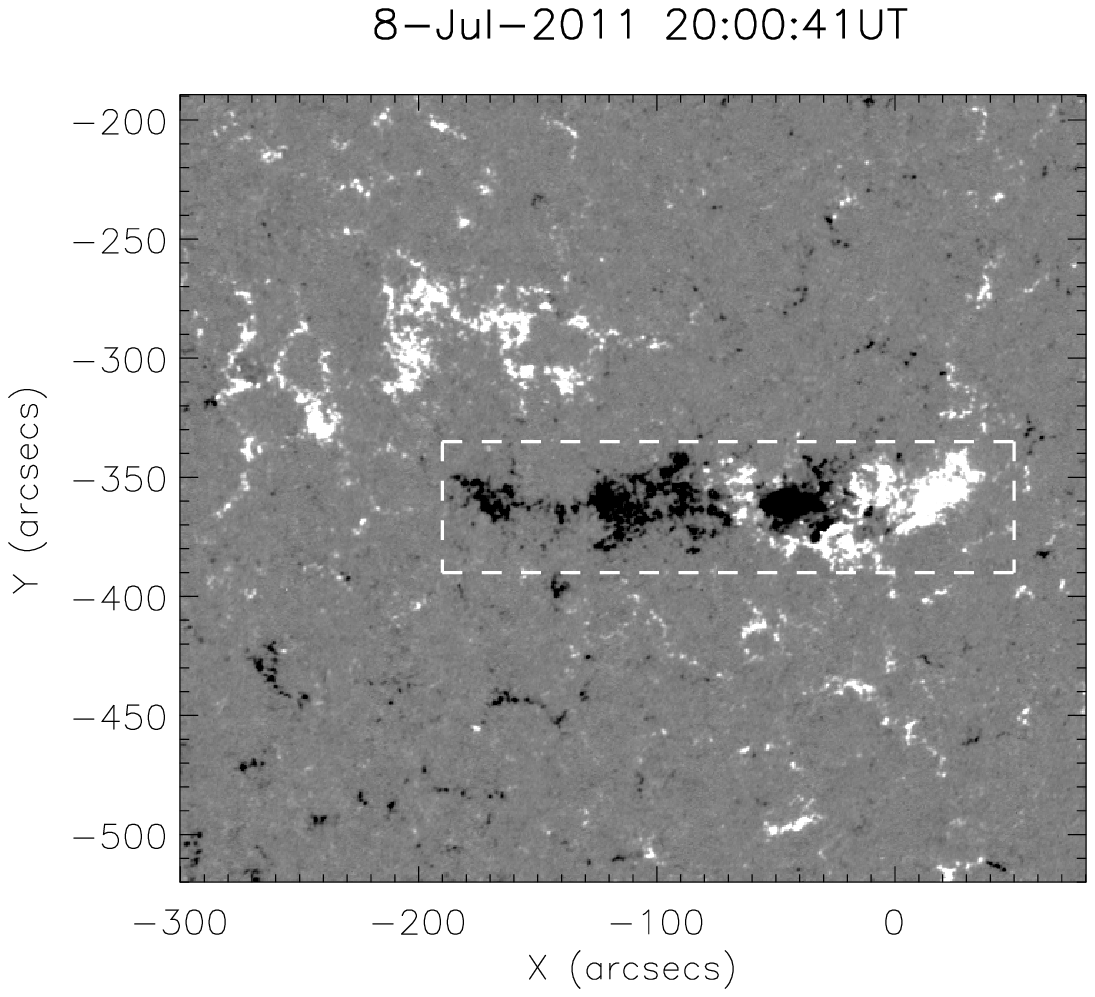}\hspace{-0.3in}\includegraphics[width=0.5\textwidth,clip=]{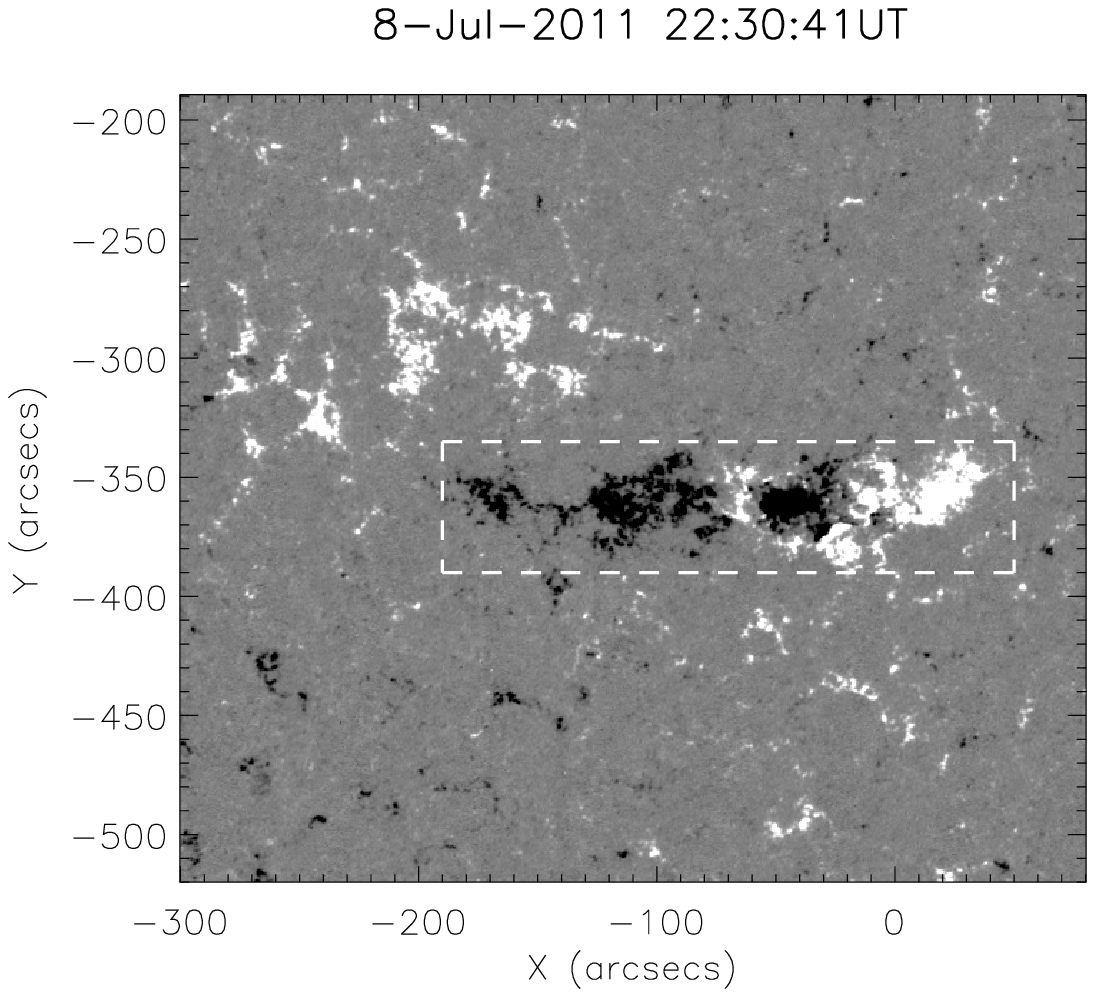} \\
\end{center}
\caption{A time sequence of magnetograms. Black and
white colors represent the negative and positive polarity of the magnetic field, respectively. The boxed region in the last
three images show the quadrupolar configuration of magnetic flux distribution.}
\label{fig:6}
\end{figure}

\subsection{Emerging flux region}

The location of the filament in the photosphere can be identified by overlying
contour of the filament extracted from 304~\AA~image on the
magnetogram. Figure~\ref{fig:5} shows the contour map of the filament overlaid
upon the magnetogram. The contour of the filament was extracted from the 304~\AA~image
obtained at 23:20~UT on July 08, 2011 which is about 5~min before the magnetogram displayed in Fig.~\ref{fig:5}.
The East side footpoint of the filament is located in the positive
polarity plage region and the West side is anchored in the negative polarity plage region.
These are shown in Fig.~\ref{fig:5} in boxed regions A and B respectively.
The box C shows the emerging flux region with positive and negative polarity flux associated with two small sunspots that were present during the eruption.

Figure~\ref{fig:6} shows a sequence of magnetograms taken at different time in
which the emerging flux region (region C) was visible. At about 00:50~UT on July 08, small mixed
polarity regions resembling salt and pepper emerged between the positive polarity regions.
At about 04:00~UT new negative magnetic flux started to emerge in the positive magnetic
field region of the Western plage.  In the meanwhile, at around 15:50~UT, an intrusion
of positive flux is observed between the pre-existing negative flux region of plage and the
newly emerged negative flux. The flux emergence and intrusion resulted in a final
configuration of quadrupolar flux distribution of the Western magnetic field region. The boxed region (shown by dashed lines) in
the last three panels of Fig.~\ref{fig:6} shows the intrusion of the positive flux into the
 negative flux region. The boxed region also shows the quadrupolar configuration of the
magnetic field distribution.

\begin{figure}
\begin{center}
\includegraphics[width=0.8\textwidth,clip=]{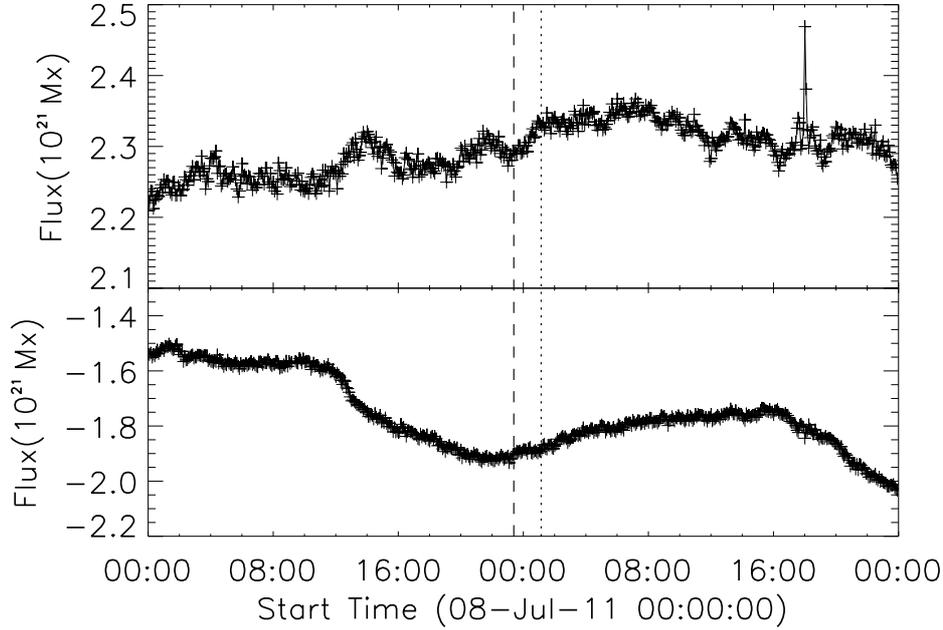}
\end{center}
\caption{Top: The evolution of flux plotted as a function of time shown for the positive
polarity plage region of Fig.~\ref{fig:5} (box A). Bottom: Same as top side plot but for negative polarity plage region of Fig.~\ref{fig:5} (box B). The first dashed vertical line represents the filament eruption time and the second dotted line
represents the starting time of the flare.}
\label{fig:7}
\end{figure}

Figure ~\ref{fig:7} shows the evolution of magnetic flux for locations A and B of the plage regions.
The flux has been computed for those pixels whose absolute magnetic field strength
is larger than 10~G in the boxed region A and B.
The positive flux in the boxed region A was almost constant
until about 08:00~UT on July 08, 2011. After 08:00~UT it increased slowly
for few hour and became little faster after the eruption 
(Fig.~\ref{fig:7} (top)). After the B-class flare, the magnetic flux in this region became
almost constant for about 5~h. The filament footpoint associated with this region were detached
in the later part of eruption. The negative flux in Western footpoint of the filament
(shown as box B) exhibited a different type of flux evolution. The flux started to increase
from the beginning of the observations (Fig.~\ref{fig:7} (bottom)). This was because of a
small negative flux region emerging in the vicinity of the Western plage.
The increase in flux is also due to the merging of separated regions in plage, thereby
increasing the strength of the plage region, though the area decreased slightly. The increase
in flux was small in the beginning of Jul 08, 2011. Later, at around 12:00~UT onward the
negative flux started to increase until an hour before the filament eruption. Western footpoint
of the filament associated with this region had erupted at around 23:20~UT. The flux in
the associated region had started to decrease about an hour before the filament eruption.

\begin{figure}
\begin{center}
\includegraphics[width=1.0\textwidth,clip=]{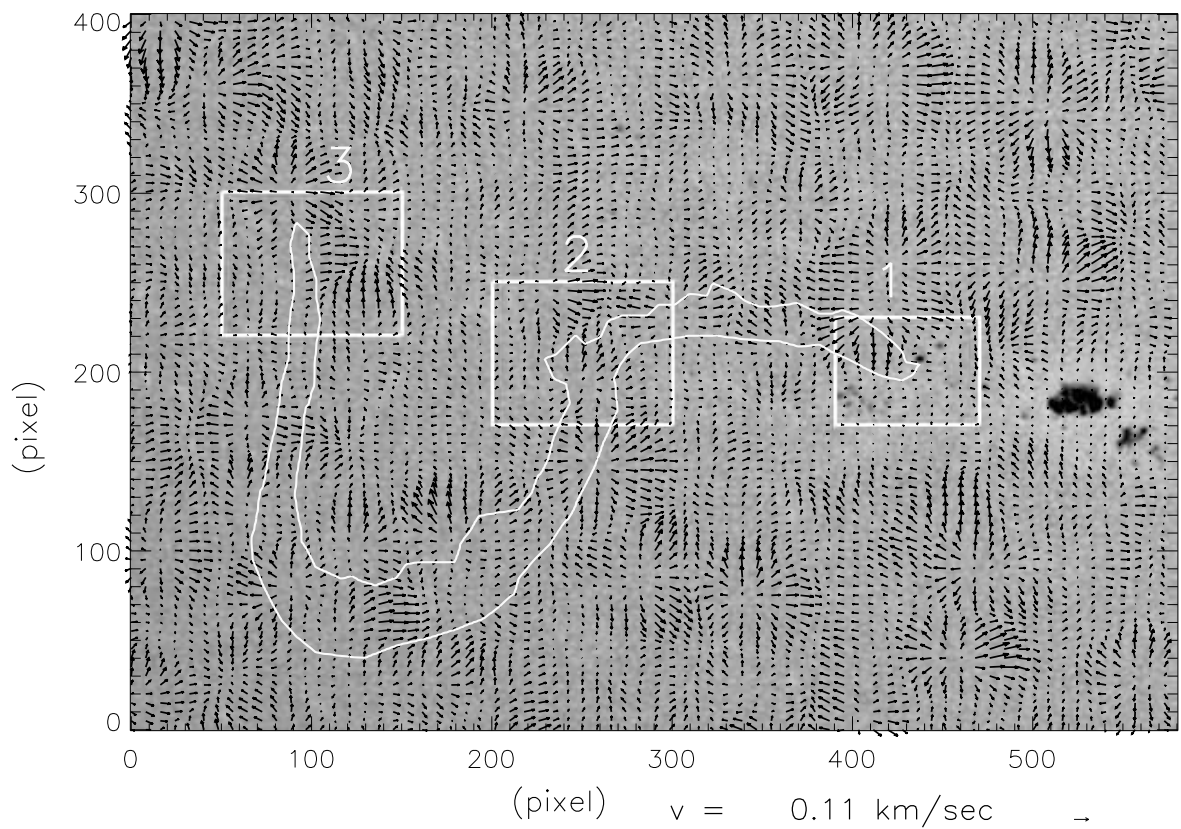}
\end{center}
\caption{The horizontal velocity vectors shown in black arrows are overlaid upon the
continuum image. The boxed region 1 and 2 show Western footpoint of the filament and the
location of the bifurcation. The boxed region 3 shows the Eastern footpoint of the filament.
The axes are labeled in pixel units.}
\label{fig:8}
\end{figure}

\subsection{Flows in and around the filament footpoints}

Figure~\ref{fig:8} shows the photospheric flow in and around the sunspot regions,
plages and filament. The flow field has been obtained by using the local correlation
tracking \citep[LCT;][]{November88} technique applied on continuum images. The images
used for obtaining the horizontal velocity of the photospheric plasma
are three minute apart and the Gaussian apodizing window function is 4.5$^{\prime\prime}$.
The obtained velocity vectors were integrated over two hours period. The results mainly
shows the long lived flows in and around the sunspots and plage regions. In the sunspot
group an outward flow has been observed. Besides, a long lived flows in the quiet sun is a
diverging motions whose size is about 30-40 arcsec. The pattern is similar to the one observed by
\cite{derosa04}. These diverging flow patterns are the supergranular outflows that are
seen everywhere. The box 1 shows the location of the Western footpoint and
box 3 shows the Eastern footpoint of the filament.
The box 2 region is a location of the
another part of bifurcated footpoint in the Western side. In boxes 1 and 2 the flow direction
is always converging inward.

\begin{figure}
\begin{center}
\includegraphics[width=0.5\textwidth,clip=]{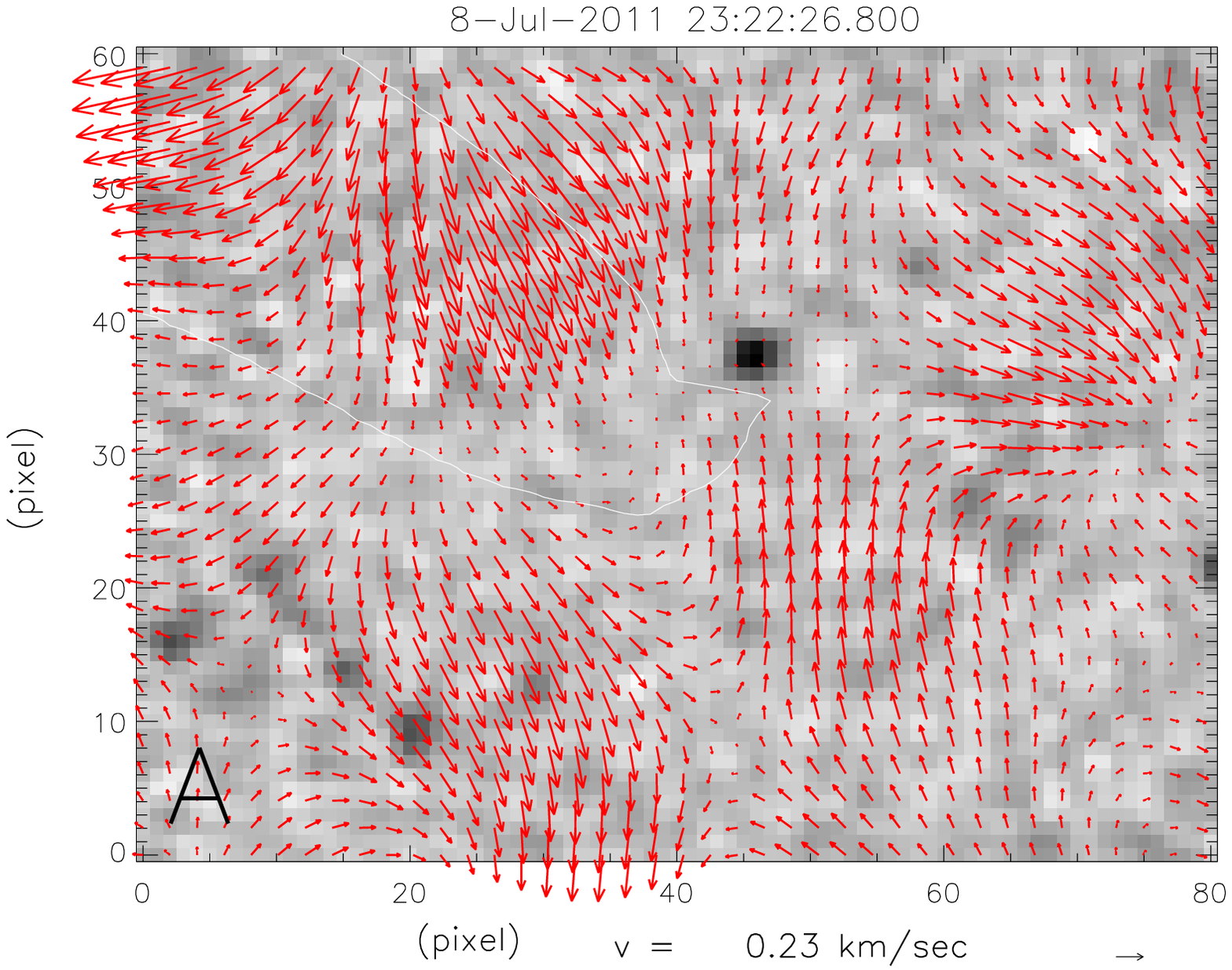}\hspace{-0.12in}\includegraphics[width=0.5\textwidth,clip=]{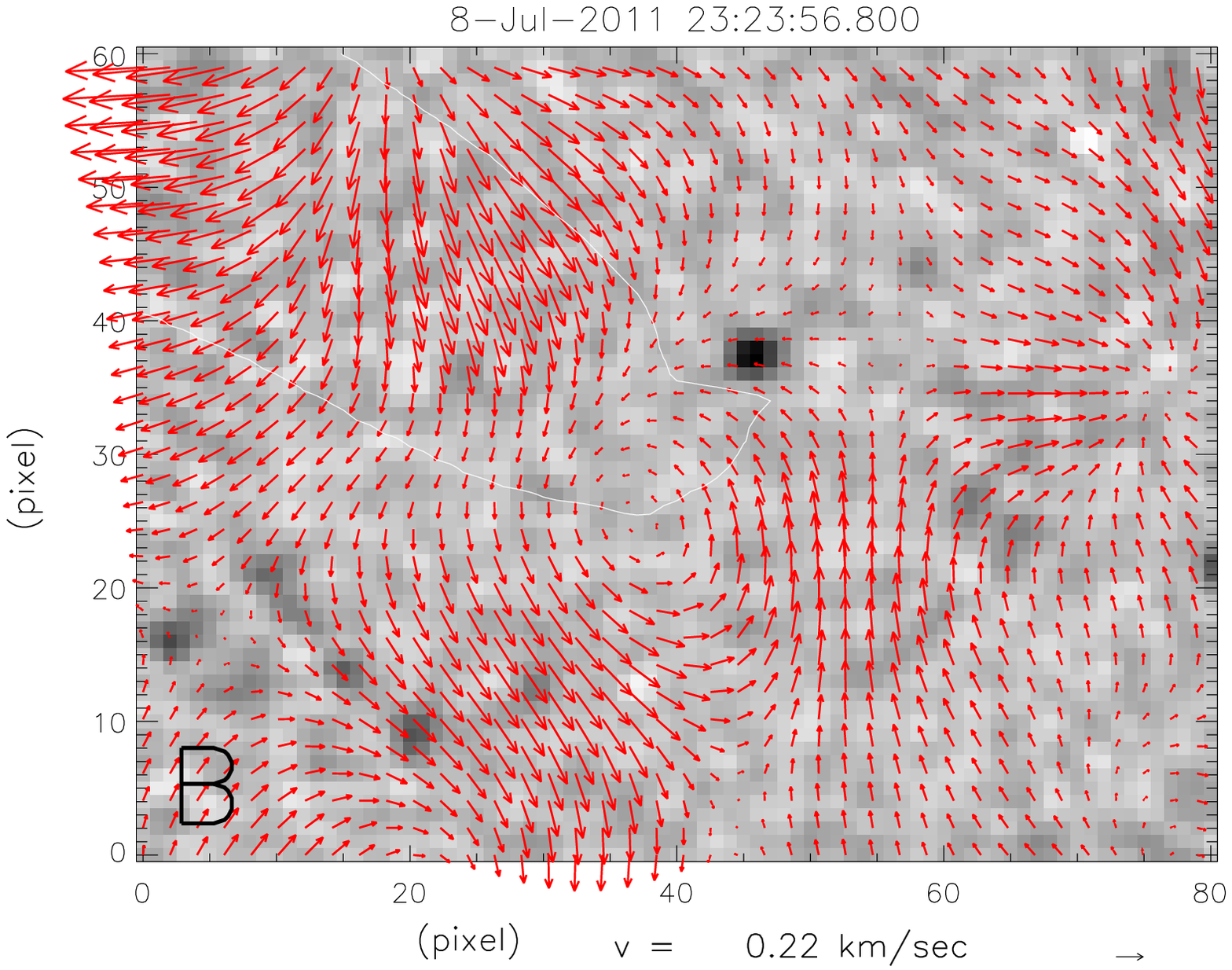} \\
\includegraphics[width=0.5\textwidth,clip=]{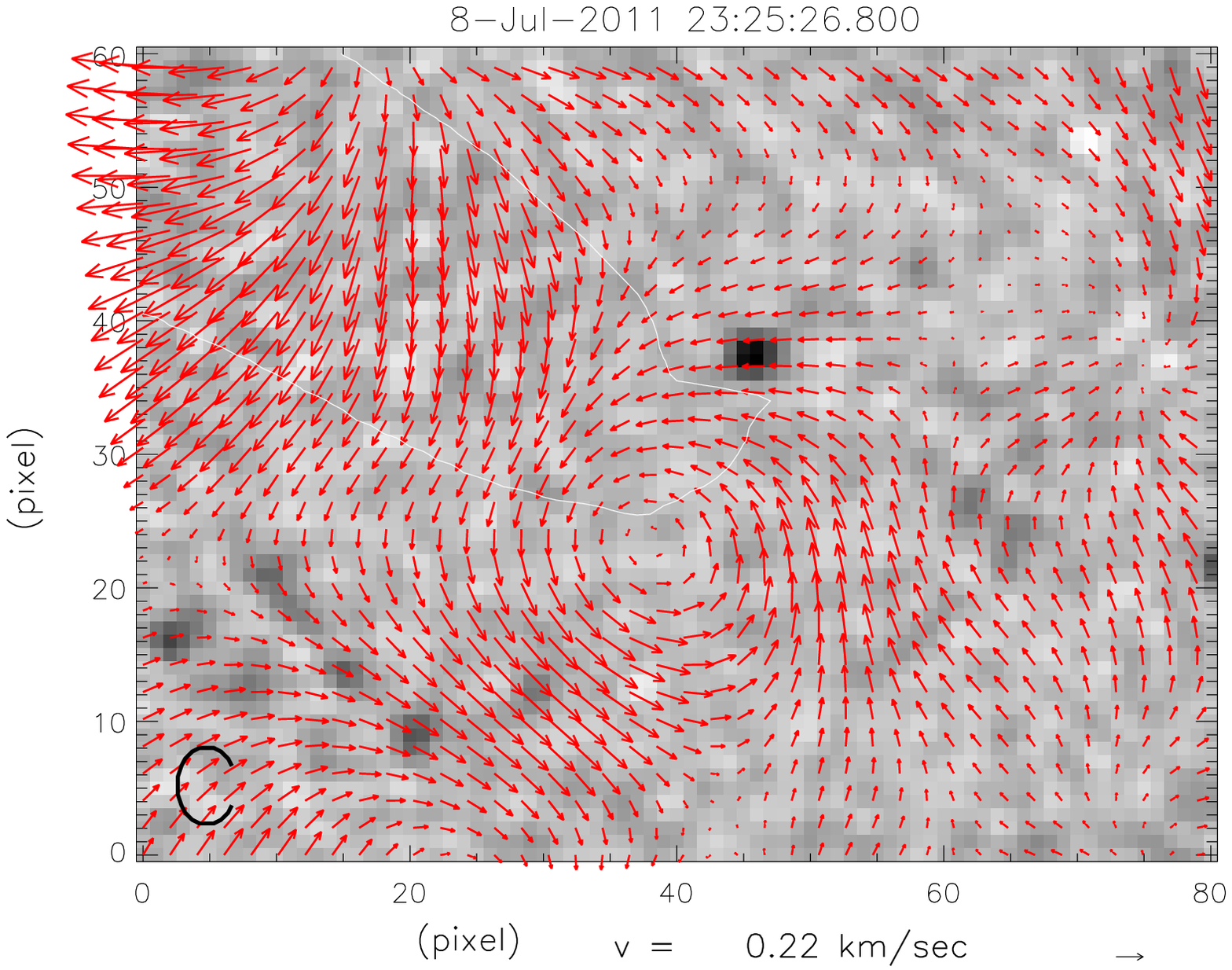}\hspace{-0.12in}\includegraphics[width=0.5\textwidth,clip=]{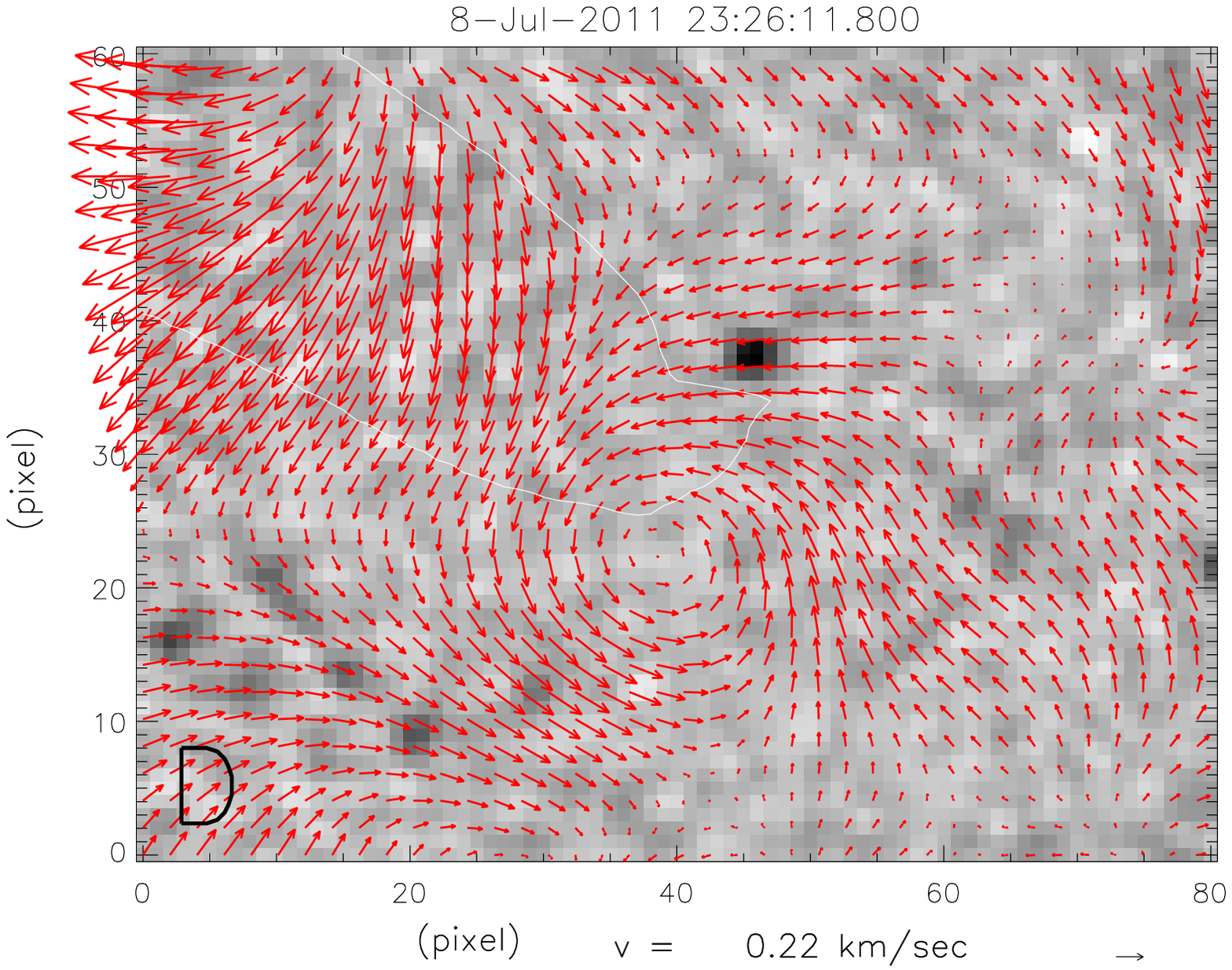} \\
\includegraphics[width=0.5\textwidth,clip=]{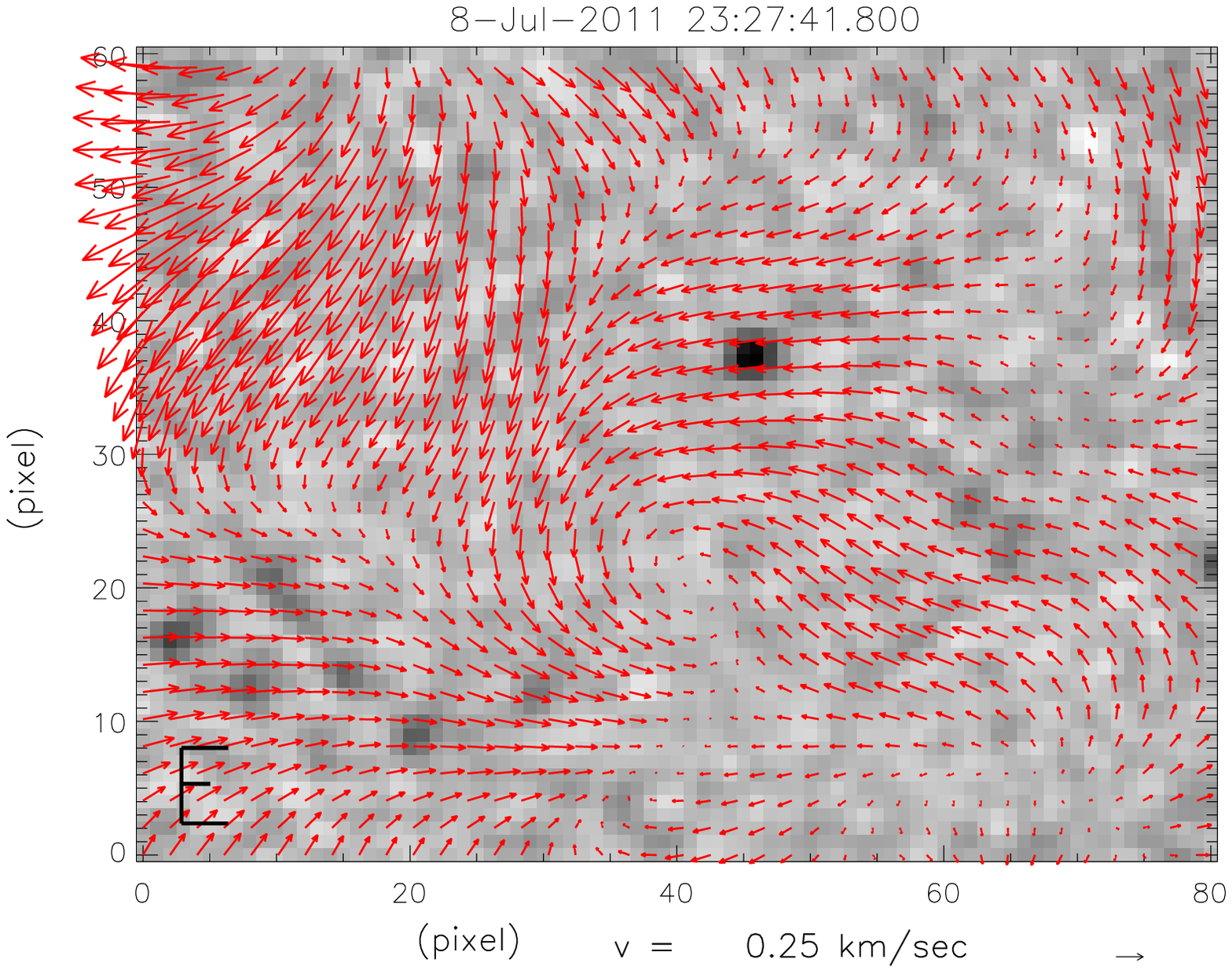}\hspace{-0.12in}\includegraphics[width=0.5\textwidth,clip=]{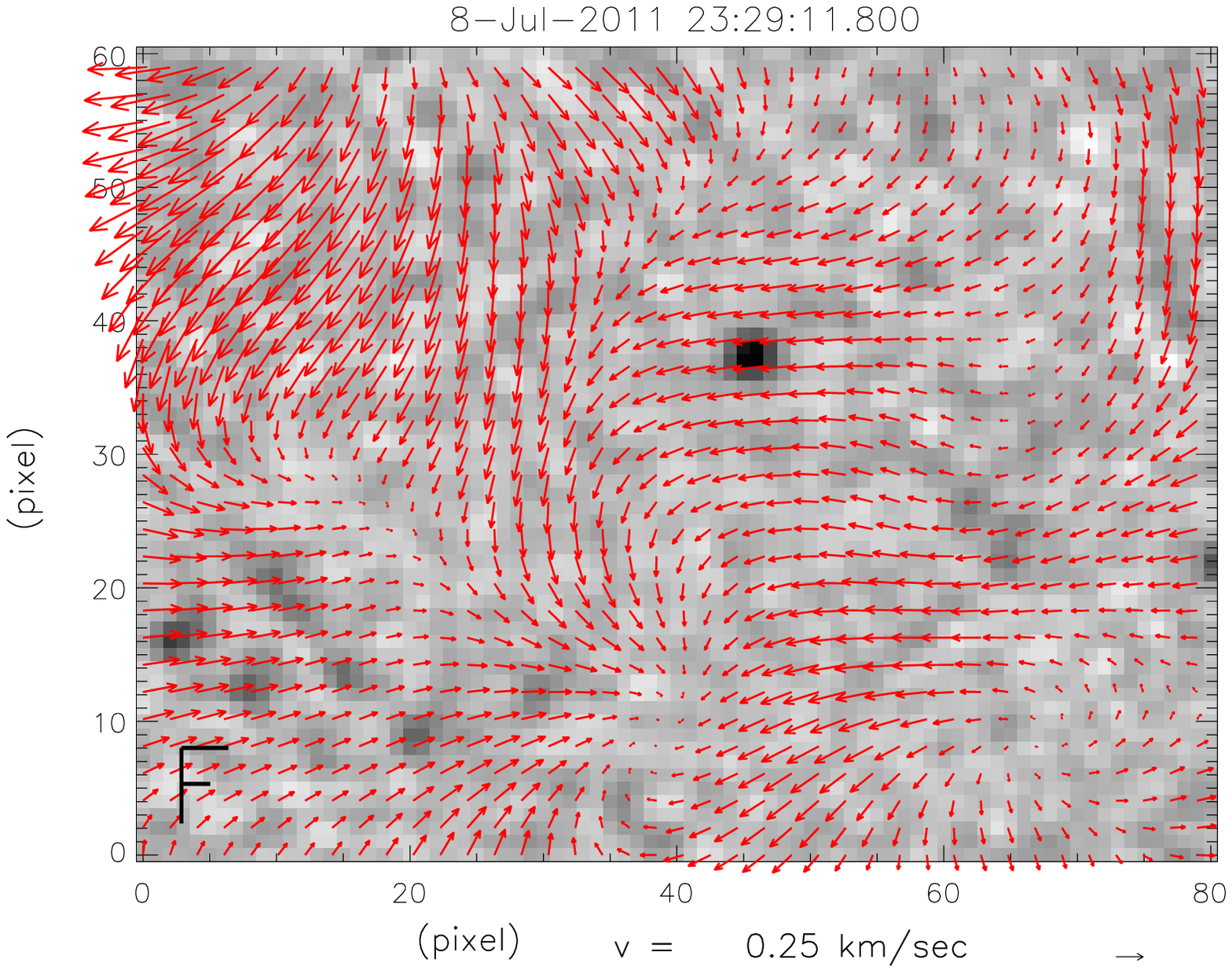} \\
\end{center}
\caption{The temporal evolution of rotational velocity pattern observed in location 1 of
Fig.~\ref{fig:8}. The date and time of the computed velocity is shown on the top of each map.
The size of arrow in the bottom of each map represents the magnitude of velocity. The horizontal
and vertical axes are shown in terms of pixels.}
\label{fig:9}
\end{figure}

The long term behavior of the flow field in the filament region is recorded in Fig.~\ref{fig:8}.
The space-time diagram shows that the filament activation initiated at around 23:20~UT.
We have also looked at the boxed regions 1 and 2 during the filament eruption time.
Figure ~\ref{fig:9} shows the temporal sequence of flow field in Western footpoint of
the filament marked 1 in Fig.~\ref{fig:8}. These velocity flow fields were obtained from
3-min time sequence images without averaging. Until 23:22~UT on Jul 08 we see the
converging flow into the small pore and filament footpoint region (Fig.~\ref{fig:9}(top-left)). However, at 23:23~UT, the
situation changed at the filament Western footpoint location and a counter clockwise rotational motion was observed. 
This  rotational motion persisted only until 23:27~UT.

\begin{figure}
\begin{center}
\includegraphics[width=0.5\textwidth,clip=]{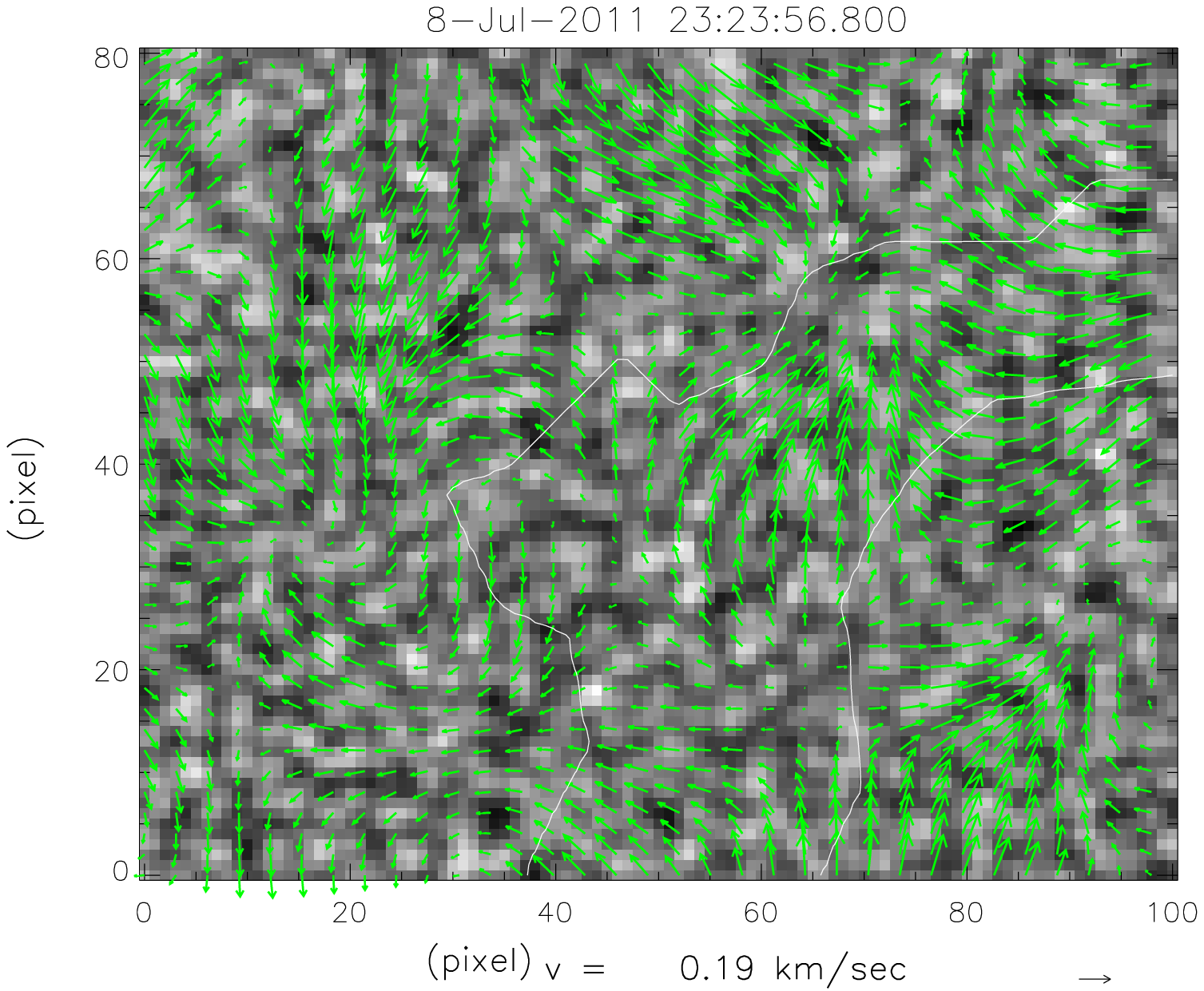}\hspace{-0.22in}\includegraphics[width=0.5\textwidth,clip=]{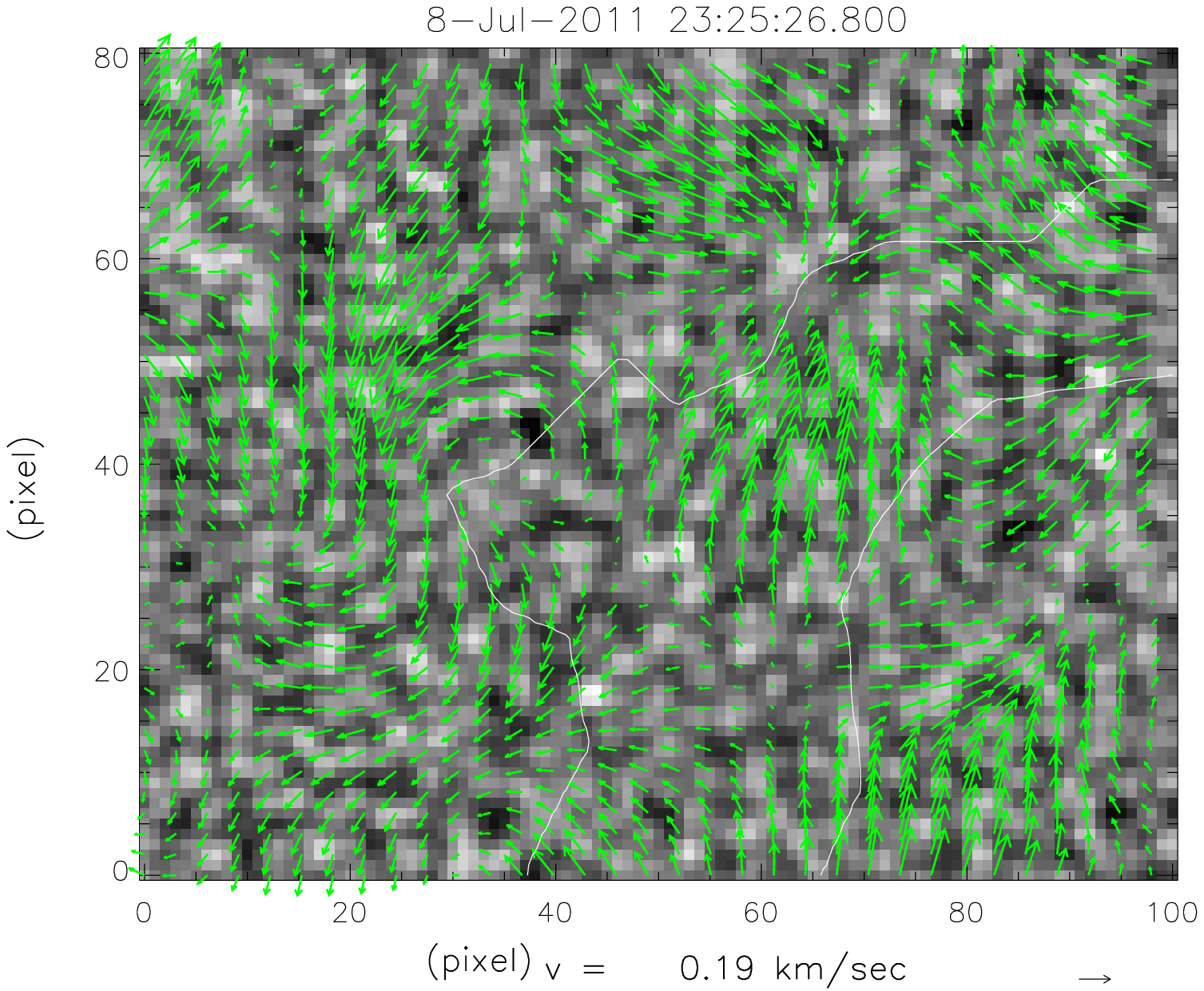} \\
\includegraphics[width=0.5\textwidth,clip=]{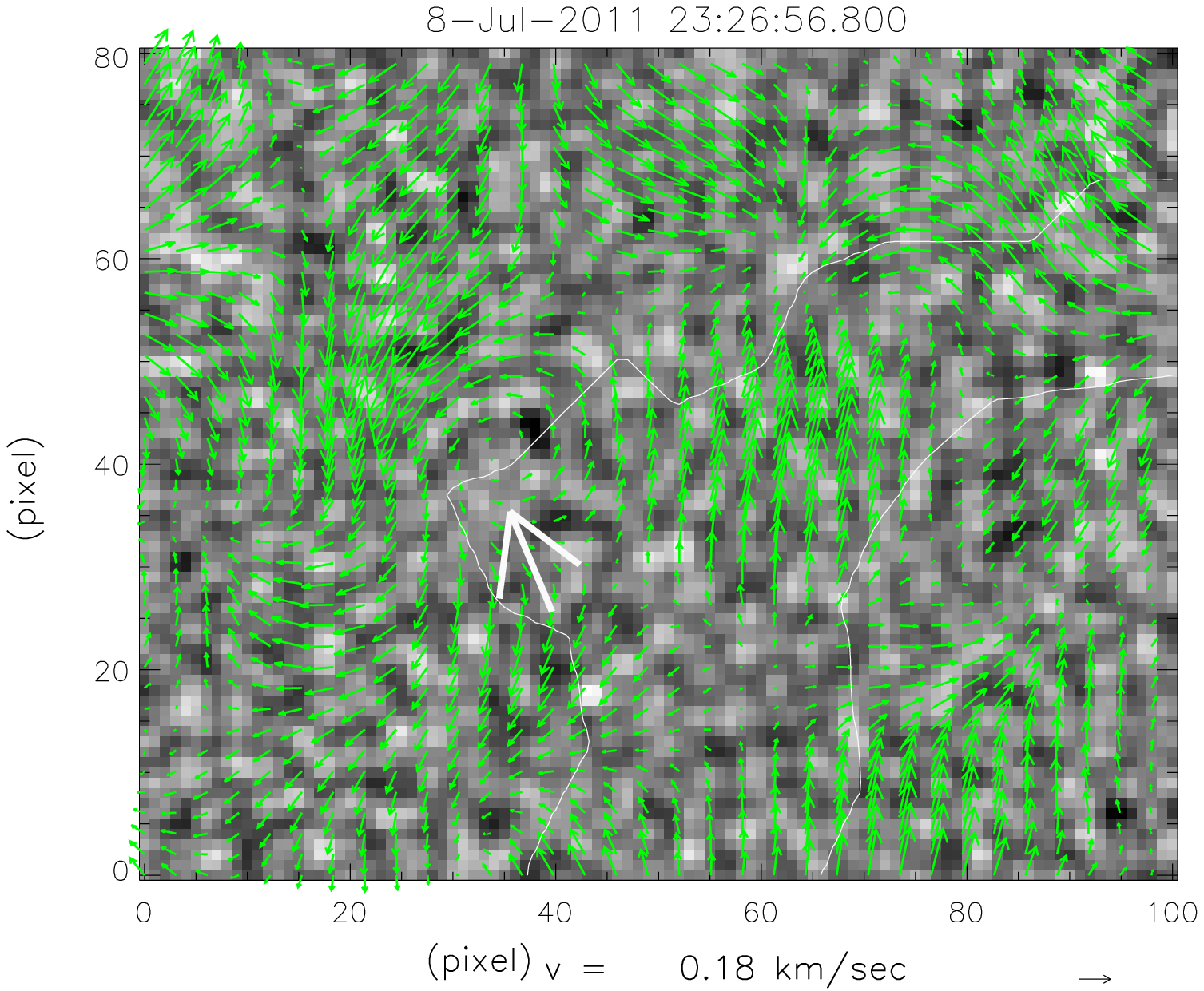}\hspace{-0.22in}\includegraphics[width=0.5\textwidth,clip=]{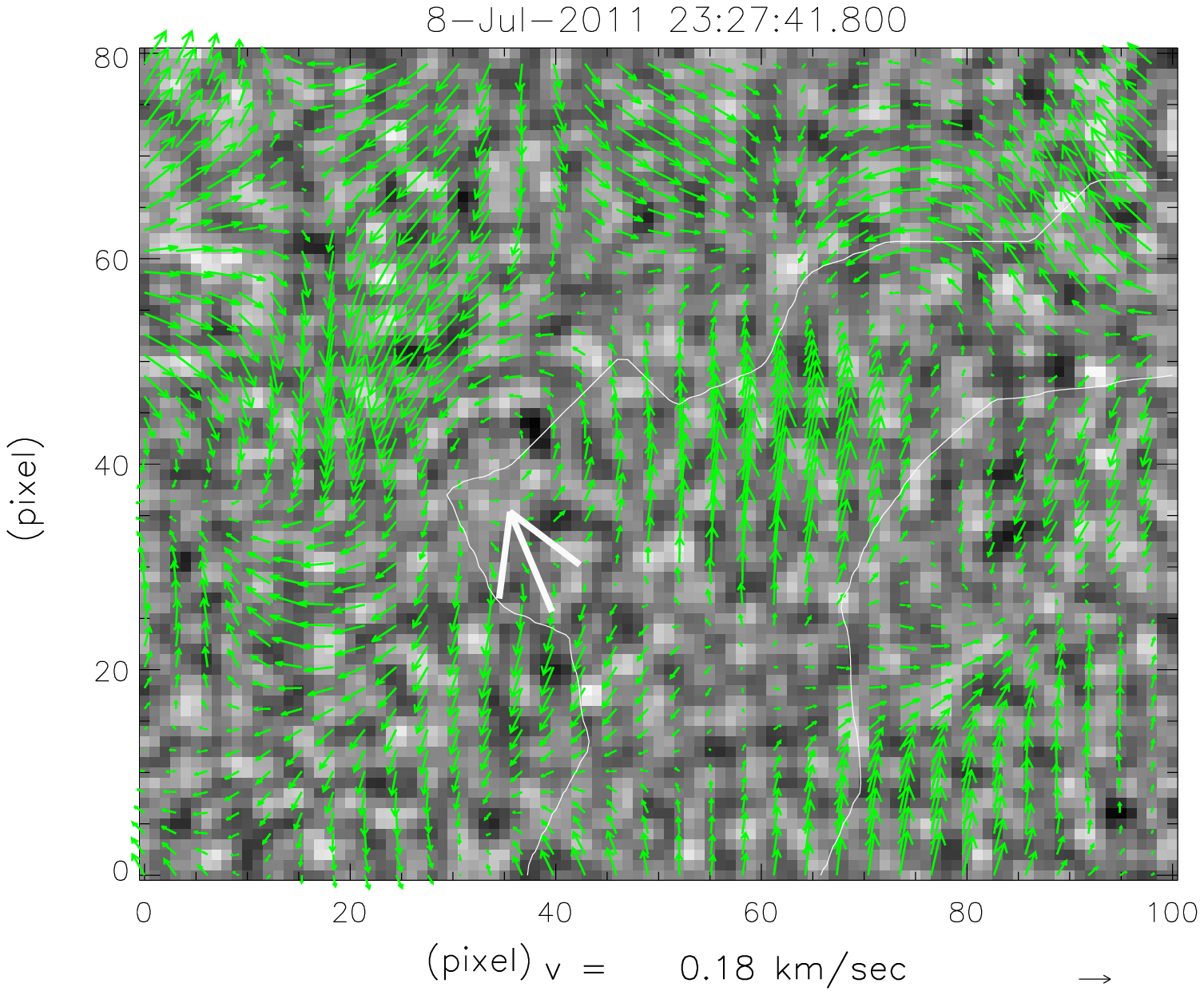} \\
\includegraphics[width=0.5\textwidth,clip=]{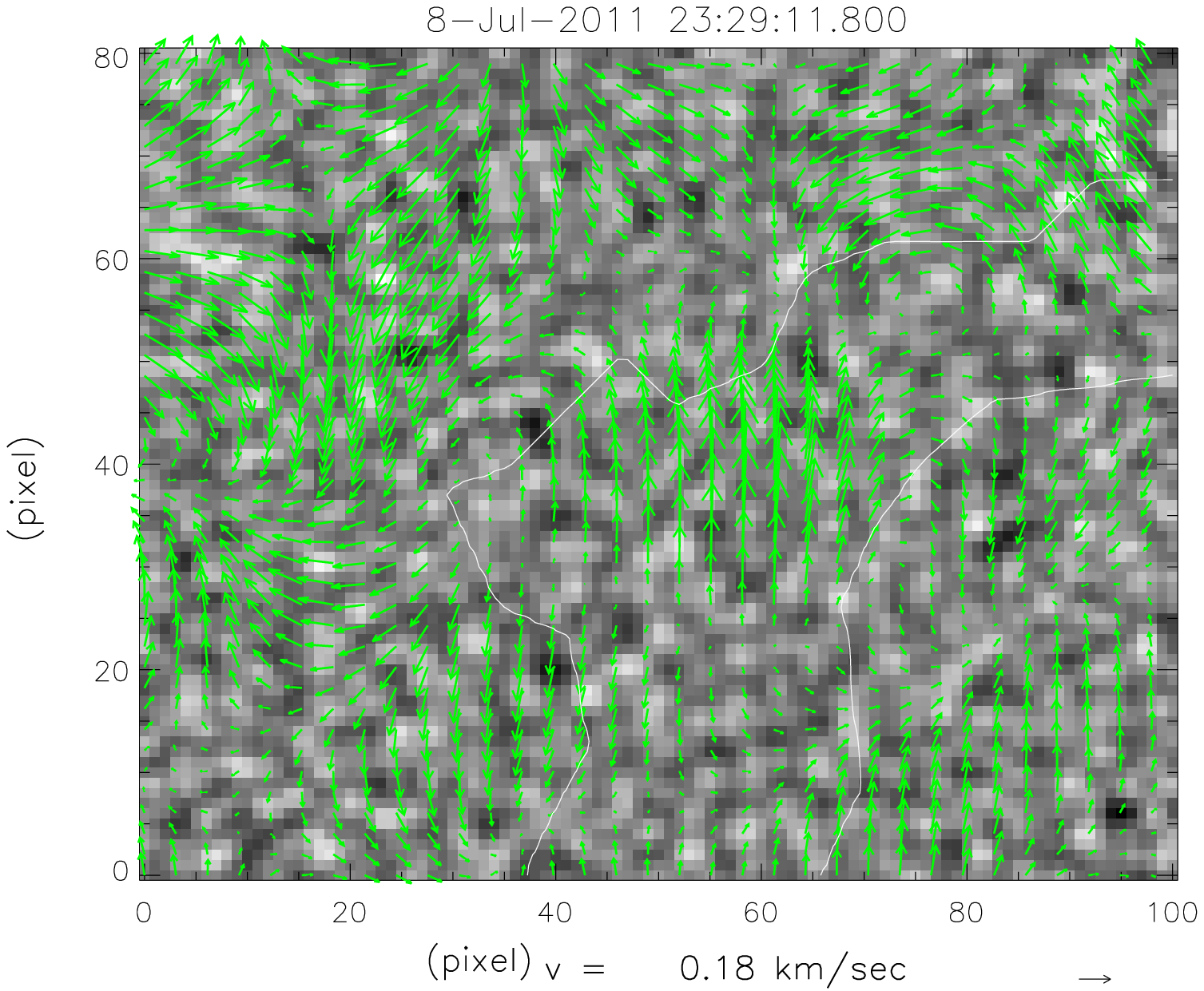}\hspace{-0.22in}\includegraphics[width=0.5\textwidth,clip=]{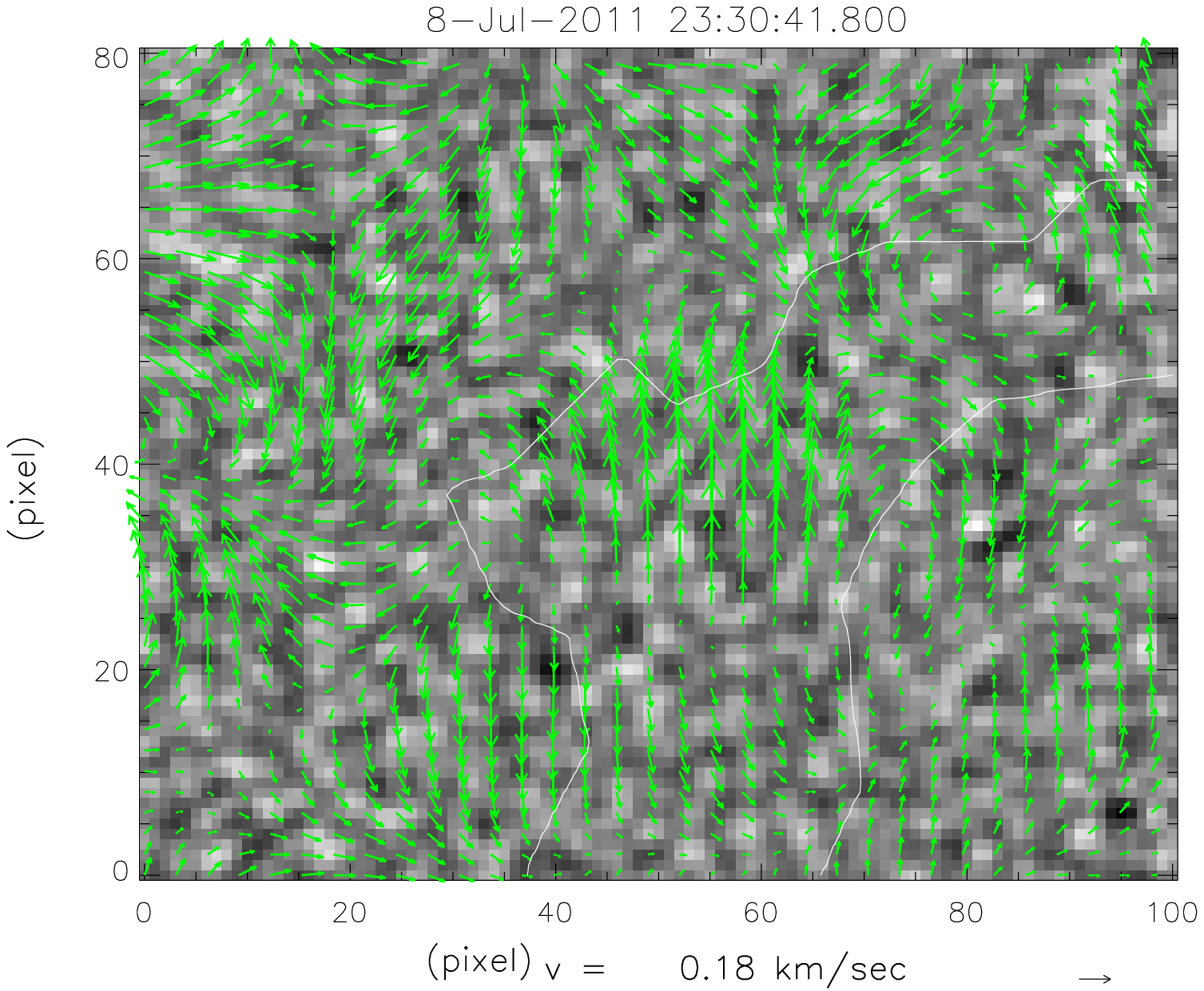} \\
\end{center}
\caption{Same as Fig.~\ref{fig:9}, but for boxed region 2 of Fig.~\ref{fig:8}. The rotation
region is shown by white arrow in the middle panels of Figure.}
\label{fig:10}
\end{figure}

Apart from location 1, the rotational motion has also been observed in location 2
(Fig.~\ref{fig:10}).
This is the another part of bifurcated footpoint of the filament in the Western portion
where the filament eruption was observed in coronal images. The location of the rotational motion is shown by the white arrow
in the middle row of Fig.~\ref{fig:10}.
This rotational motion started at around 23:24~UT and ended at around 23:29~UT.  Apart
from this the rotational motion, starting from 23:27~UT to 23:31~UT, was also observed  
in another location in top-right corner in the middle \& last row of Fig.~\ref{fig:10}.

\begin{figure}
\begin{center}
\includegraphics[width=0.5\textwidth,clip=]{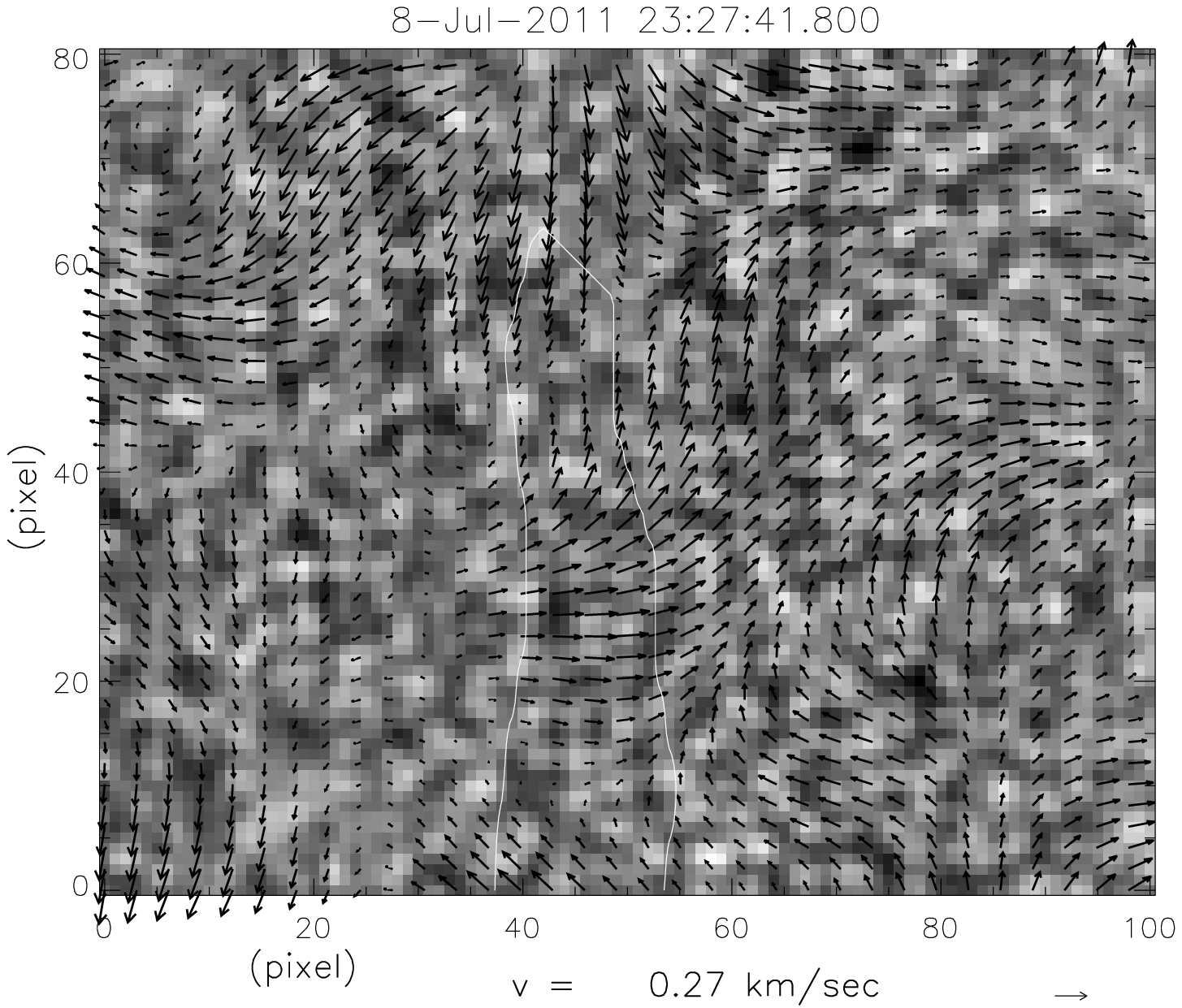}\hspace{-0.22in}\includegraphics[width=0.5\textwidth,clip=]{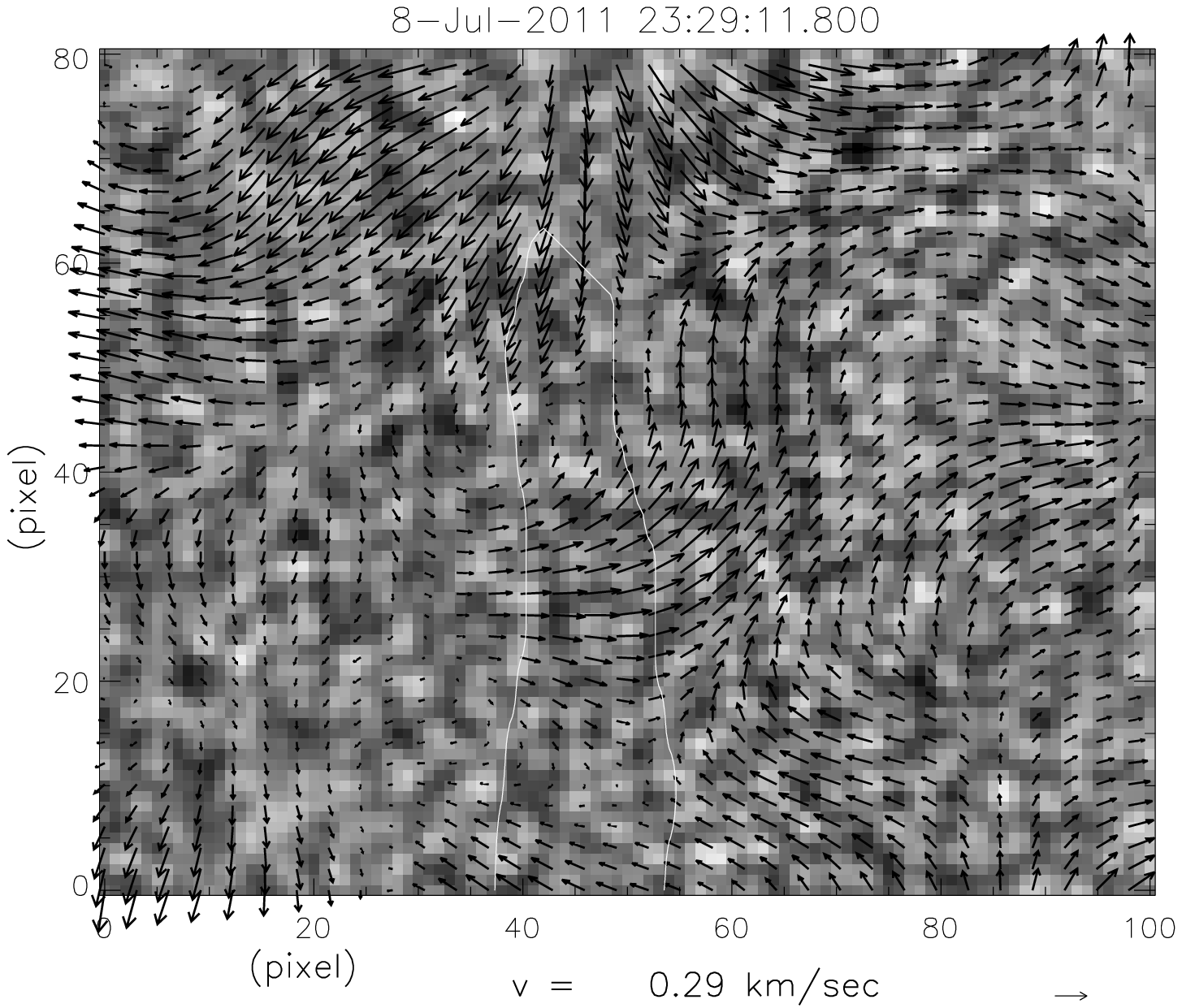} \\
\includegraphics[width=0.5\textwidth,clip=]{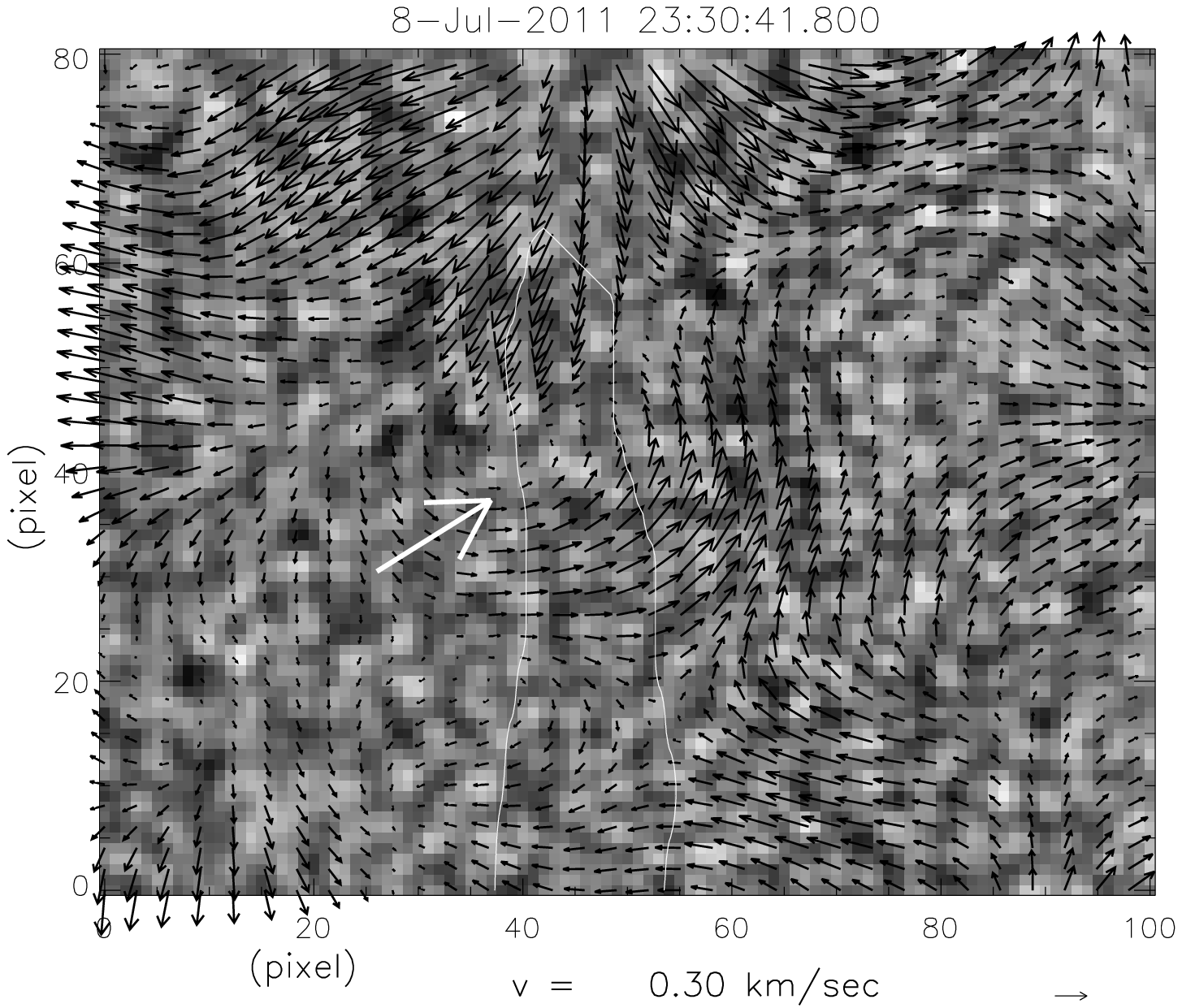}\hspace{-0.22in}\includegraphics[width=0.5\textwidth,clip=]{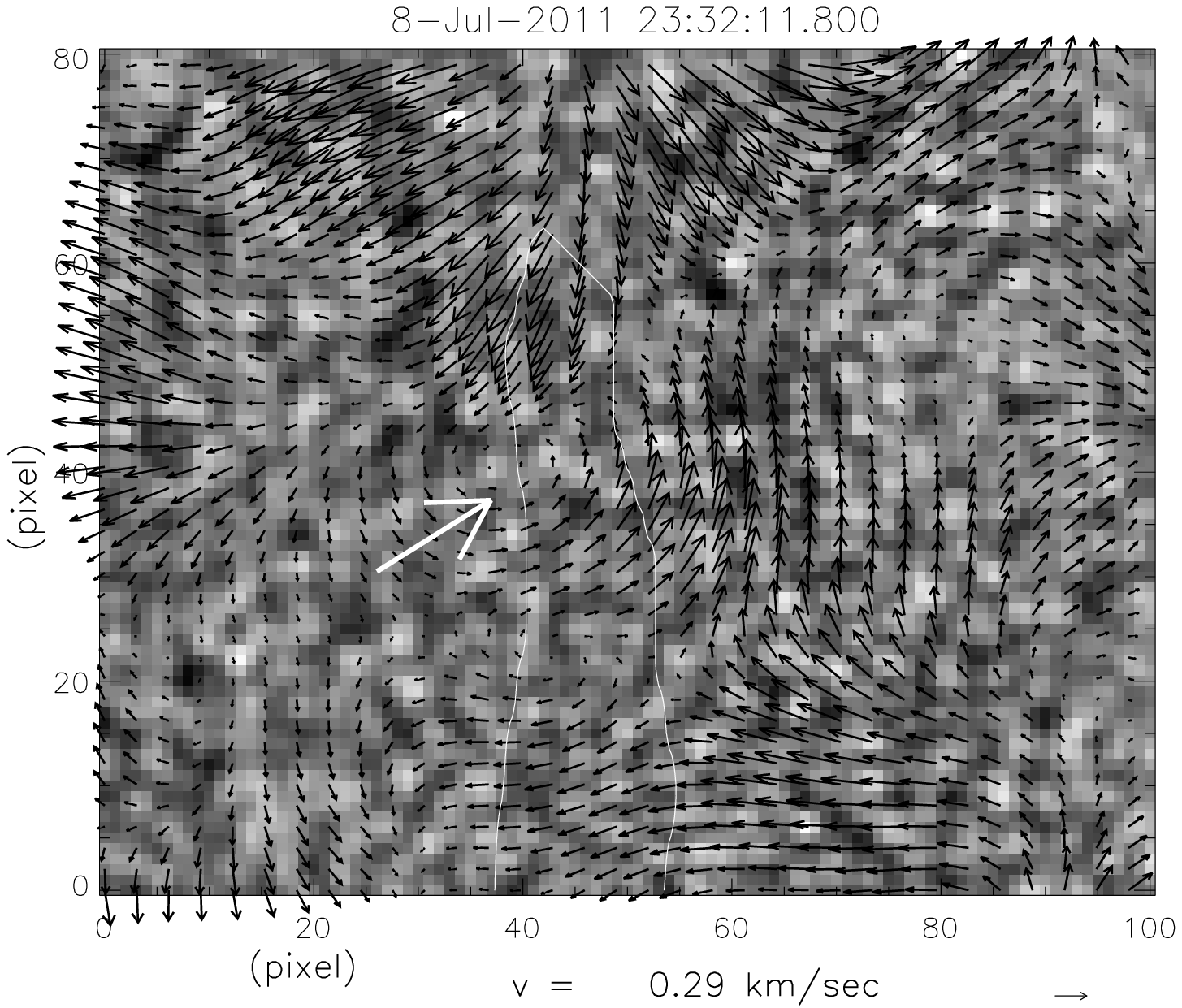} \\
\includegraphics[width=0.5\textwidth,clip=]{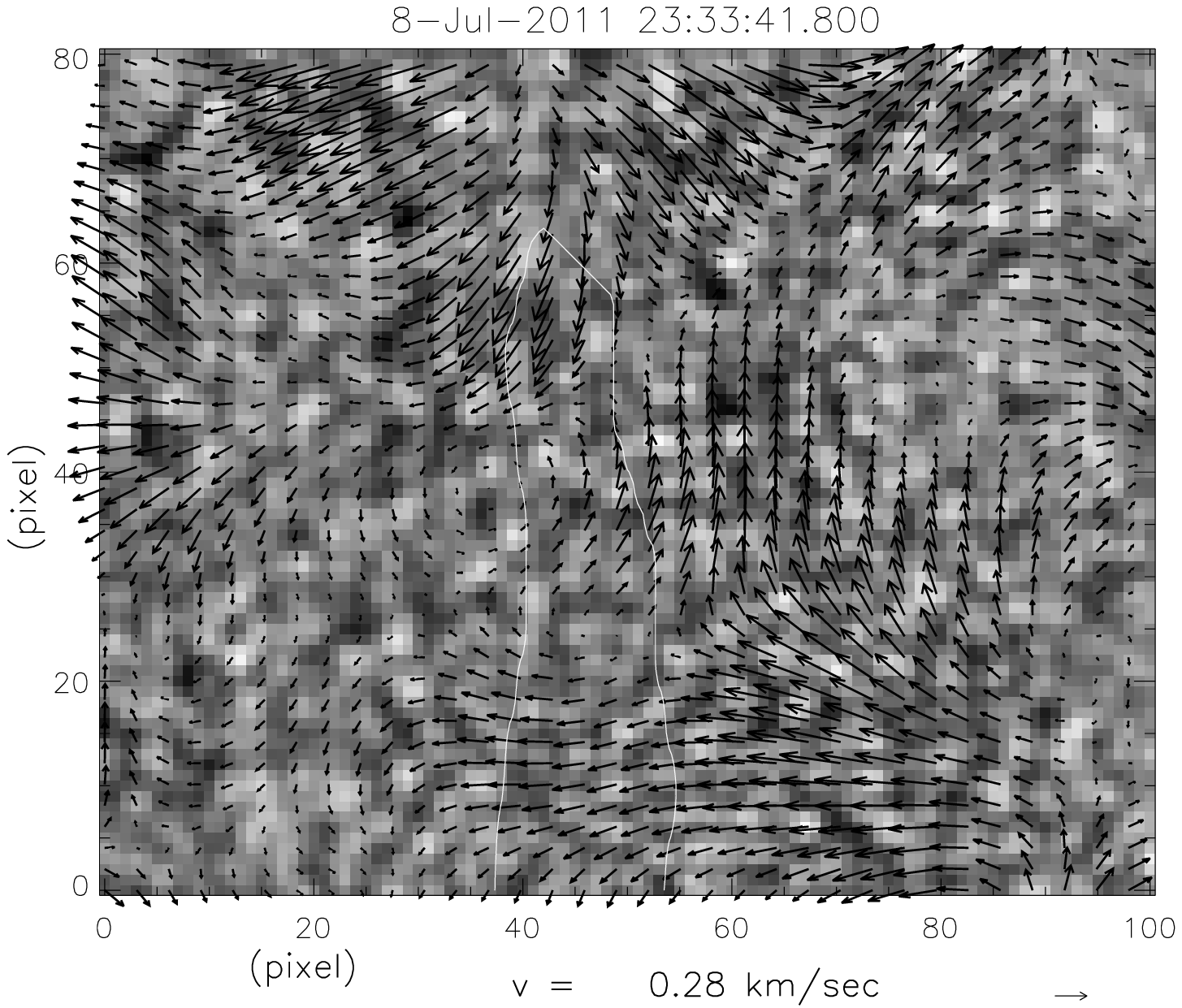}\hspace{-0.22in}\includegraphics[width=0.5\textwidth,clip=]{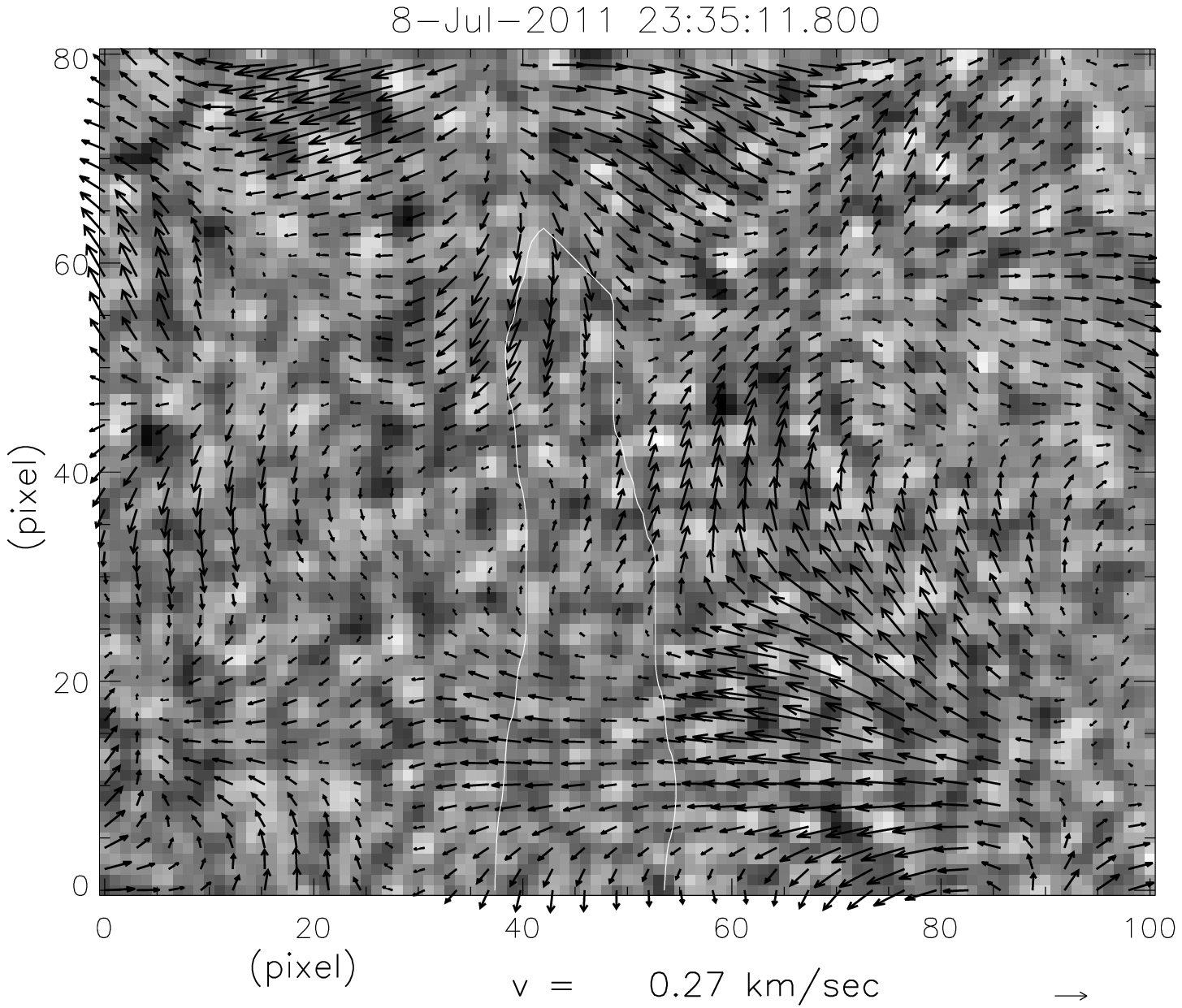} \\
\end{center}
\caption{Same as Fig.~\ref{fig:9}, but for boxed region 3 of Fig.~\ref{fig:8}. The rotation
region is shown by white arrow in the middle panels of Figure.}
\label{fig:11}
\end{figure}

In Eastern footpoint of the filament (box 3 in Fig.~\ref{fig:8}) the anti-clockwise rotational motion has been observed
at around 23:34~UT (Fig.~\ref{fig:11}). The rotational motion continued for couple of
minutes before vanishing.

\subsection{Magnetic helicity}

The magnetic helicity provides information about the degree of twist in the magnetic
flux ropes \citep{Berger84a}.  It is a conserved quantity even in reconnection processes.
However, it can vary because of changes occurring at the boundaries due to
emergence/submergence of the magnetic fields
and coronal mass ejections where the magnetic field lines opens up
partially. Following \cite{Berger84b}, it is now possible to measure the helicity injection
rate dH/dt through a surface bounding the volume as

\begin{equation}
\frac{dH}{dt} = 2\int_S{[({\bf A}_{P}\cdot{\bf B}_{h})v_{z}]}ds - 2\int_S{[({\bf A}_{P}\cdot{\bf v}_{h})B_{z}]}ds
\end{equation}
Further details and explanation of the equation is given in \cite{Ravindra08}.
Following \cite{Demoulin03}, the eq. (1) can be recast in the form,
\begin{equation}
\frac{dH}{dt} = -2\int_S{[{\bf U}_{lct}\cdot{\bf A}_{P}]{B}_{z}}dS
\end{equation}
Where, ${\bf U}_{lct} = {\bf v_{h}} - (v_{z}*{\bf B}_{h})/B_{z}$.
\cite{Pariat05} have shown that the helicity flux density measured using eq. (2) may produce some
artifacts, though the helicity flux may be correct. To overcome this problem, a modified
expression for the helicity flux density whose integration provides the helicity flux as
in eq. (2), should be used. The correct equation for the helicity flux density, therefore is,
\begin{equation}
\frac{dH}{dt} = -\frac{1}{2\pi}\int_S\int_{S^{'}}\frac{d\theta({\bf r})}{dt}B_{n}B^{'}_{n}ds ds^{'}
\end{equation}
where, $\frac{d\theta}{dt}$ represents relative rotation rate and $B_{n}B^{'}_{n}$ are
the magnetic field at two different locations.

\begin{figure}
\begin{center}
\includegraphics[width=0.8\textwidth,clip=]{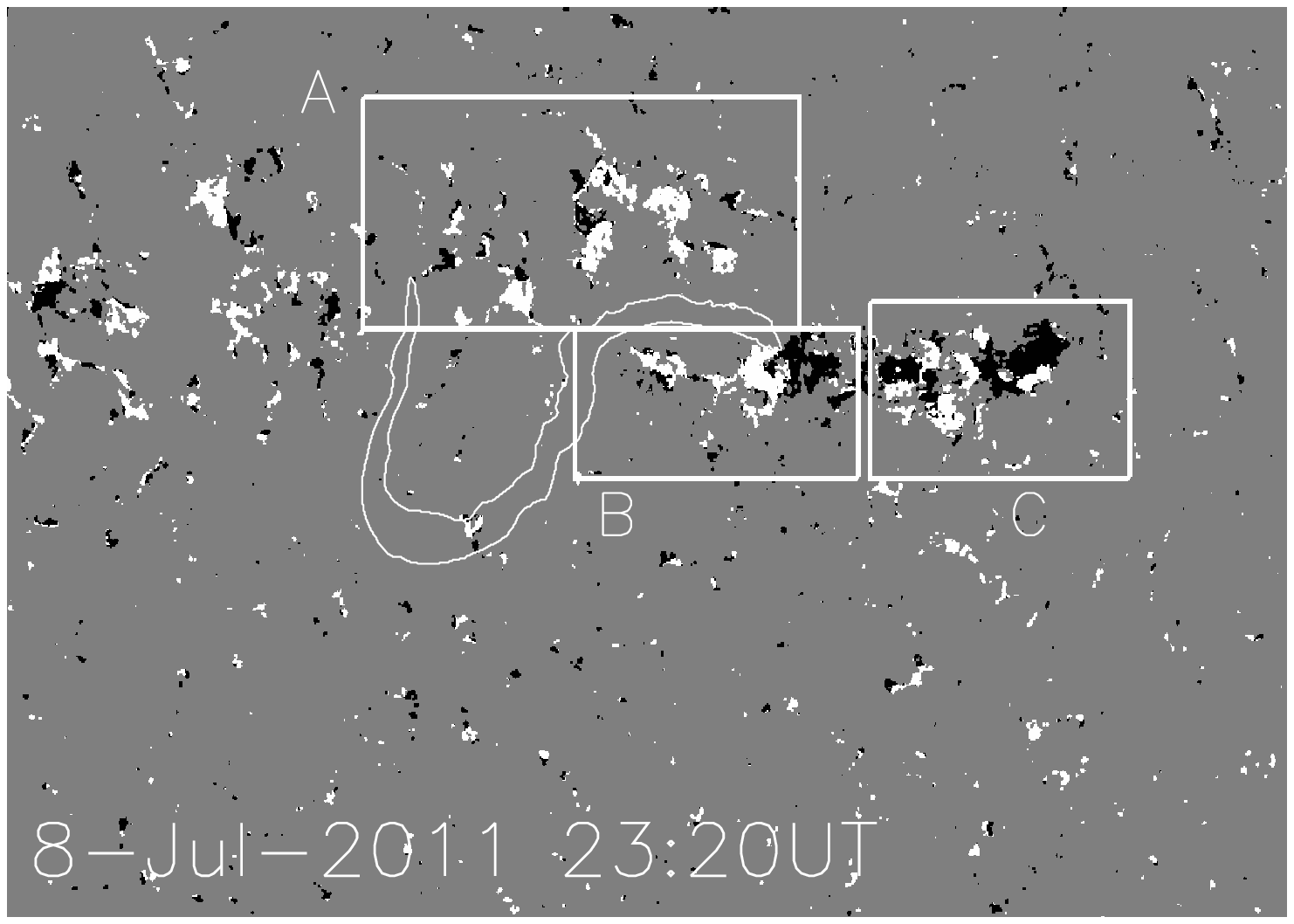} \\
\end{center}
\caption{A map of helicity flux density for the active region. The boxed region is same as in
Fig.~\ref{fig:5}.}
\label{fig:12}
\end{figure}

\begin{figure}
\begin{center}
\includegraphics[width=0.5\textwidth,clip=]{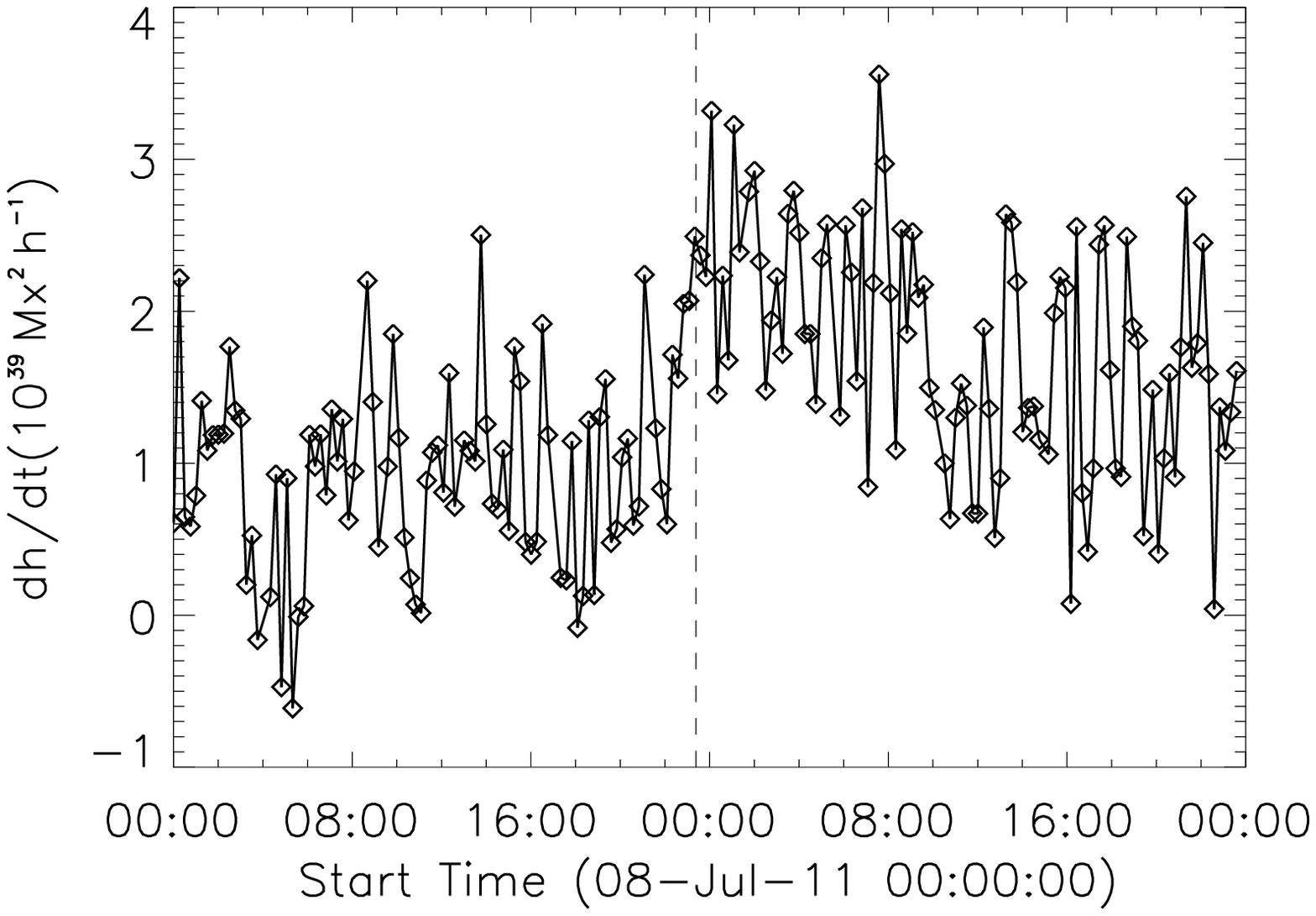}\includegraphics[width=0.5\textwidth,clip=]{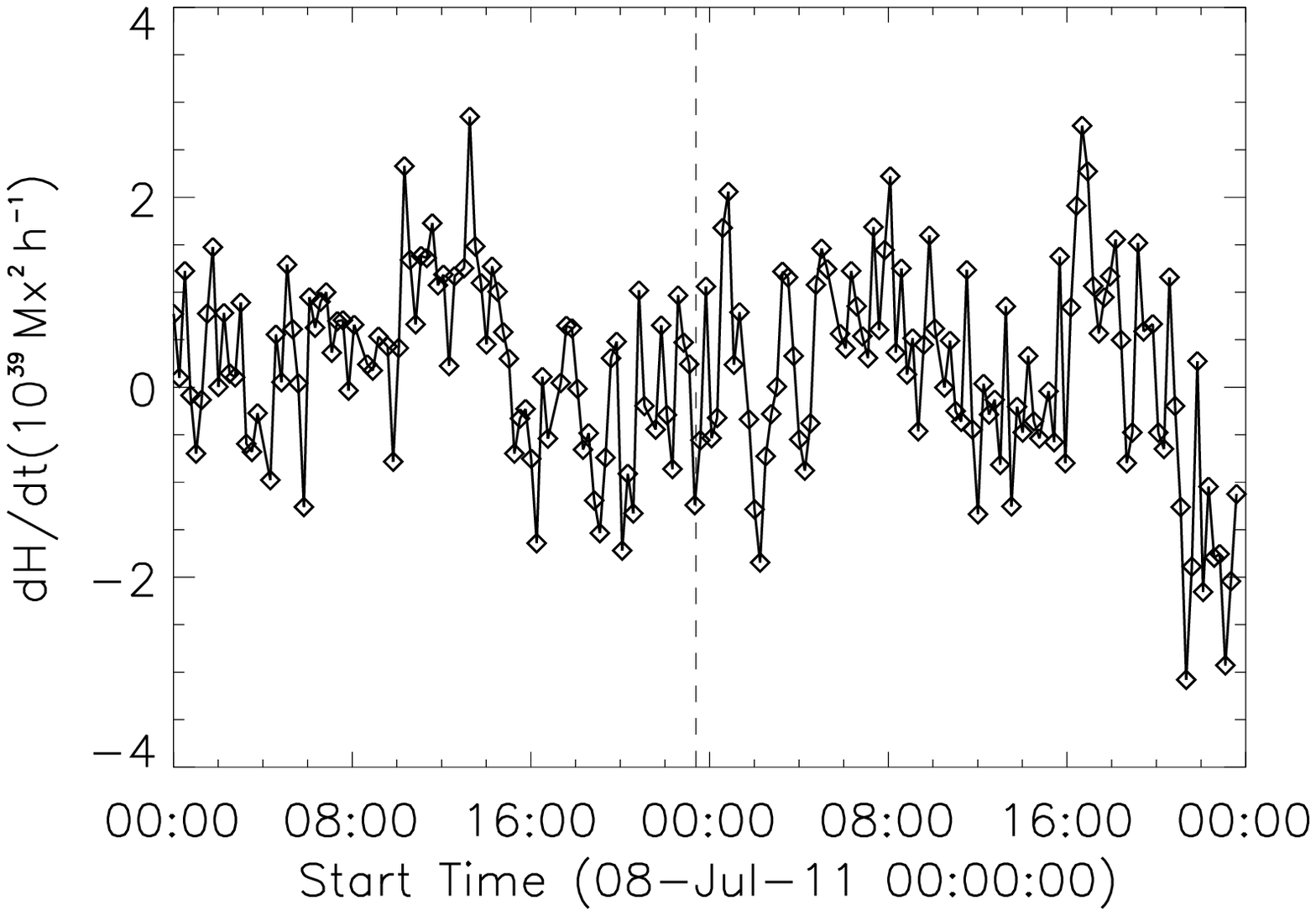}  \\
\includegraphics[width=0.5\textwidth,clip=]{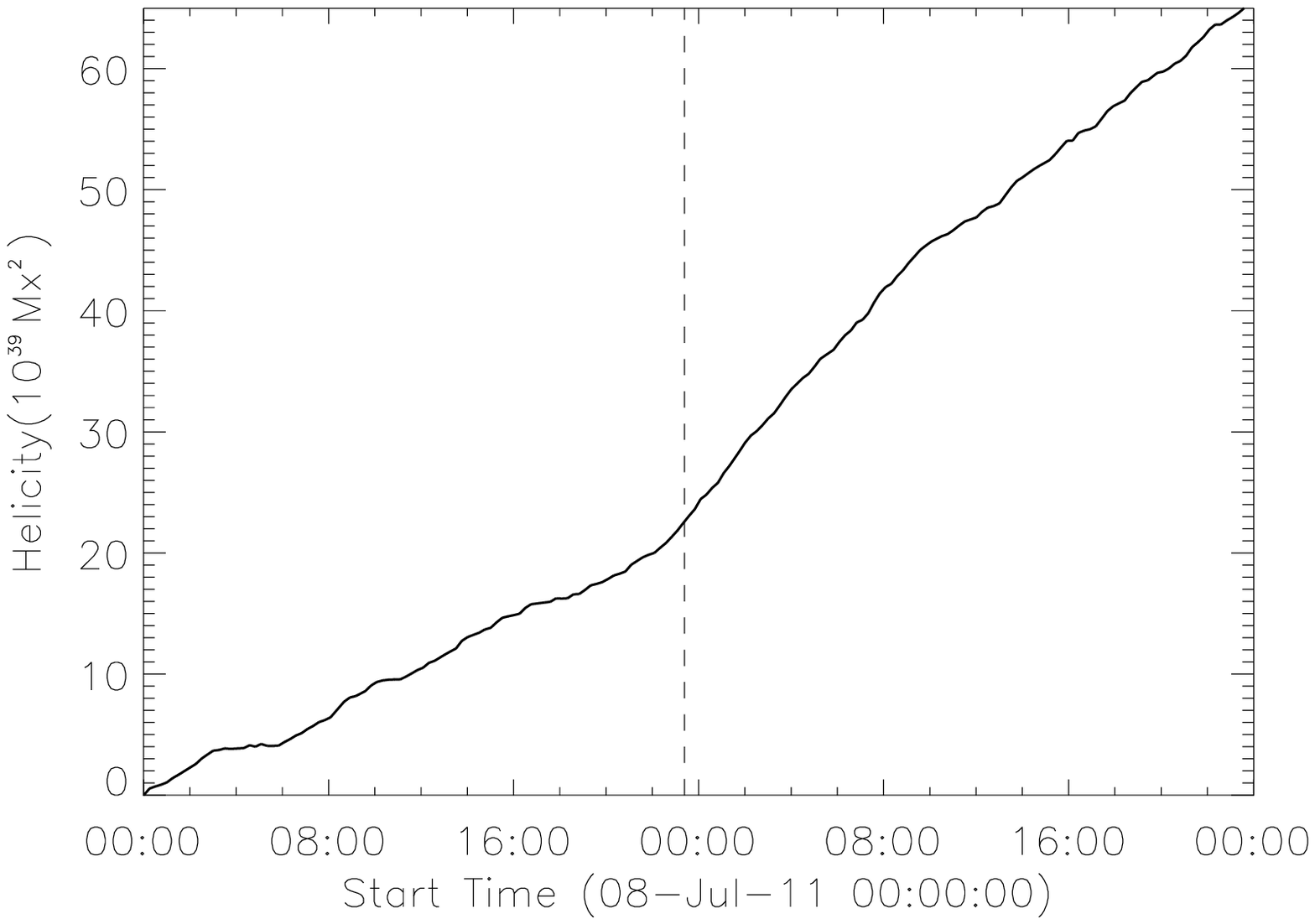}\includegraphics[width=0.5\textwidth,clip=]{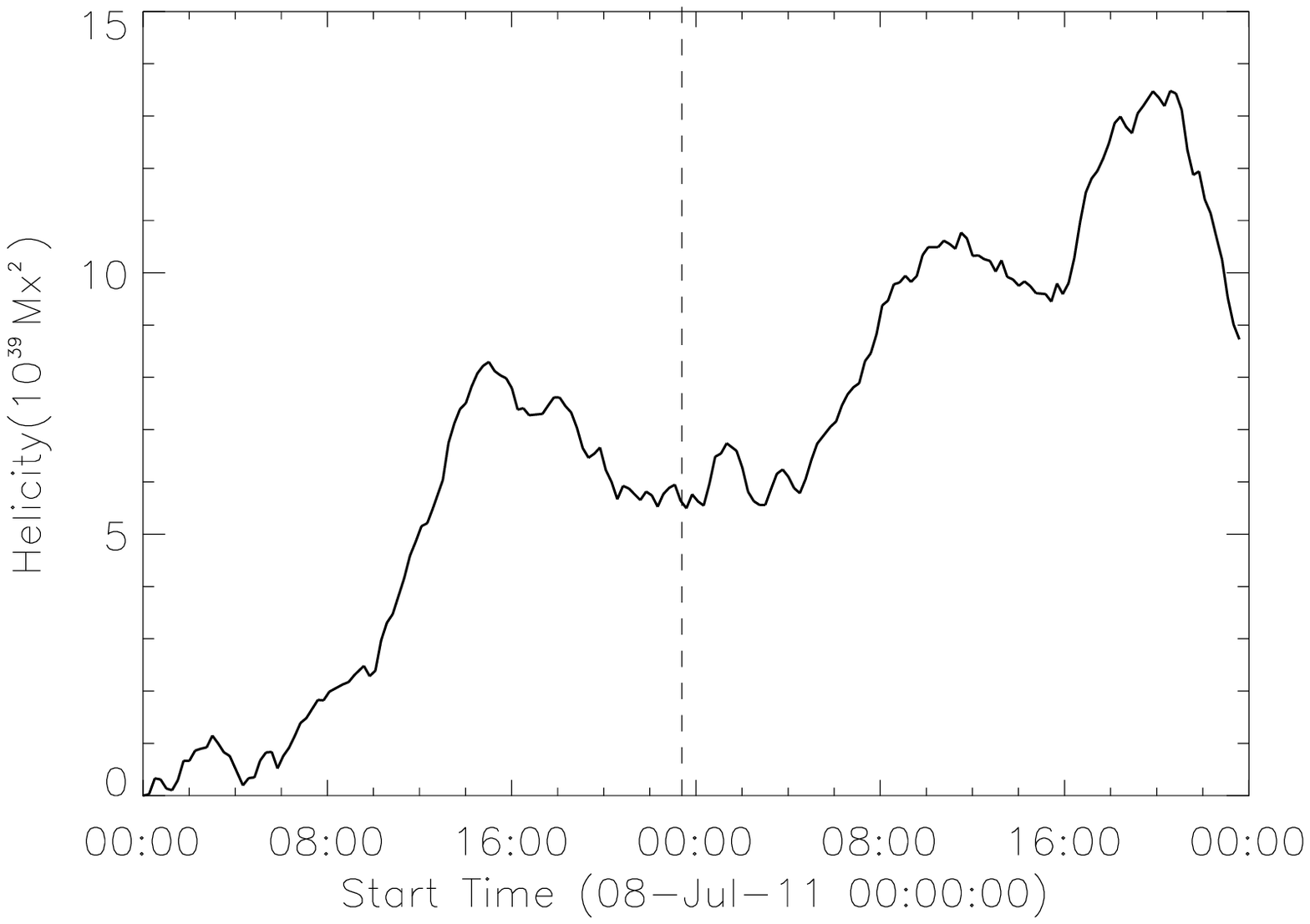}  \\
\end{center}
\caption{Top left: Evolution of helicity flux as a function of time shown for the boxed region
A of Fig.~\ref{fig:12}. Top right: Same as left side plot but for region B of Fig.~\ref{fig:12}. Bottom left: The
integrated helicity flux is plotted as a function of time for region A.
Bottom right: Same as left side plot but for region B. The dashed
vertical line represents onset time of filament eruption.}
\label{fig:13}
\end{figure}

To compute the helicity flux density we measured the magnetic footpoint velocity using
the Differential Affine Velocity Estimator \citep[DAVE;][]{Schuck05} method applied to a
sequence of magnetograms.
In any velocity detection technique it is essential to select the proper
window size and suitable time difference between
the two images.  While applying DAVE to the sequence of magnetograms, we have used a
9$^{\prime\prime}$$\times$7.2$^{\prime\prime}$ pixel box size as the apodizing
window and 15 min as
the time difference between images. It has been observed that rectangular size apodizing
window seems to work better than the square window \citep{Welsch07}. Hence, after experimenting
with different
sizes of apodizing window, we choose 9$^{\prime\prime}$$\times$7.2$^{\prime\prime}$ size in our velocity field computation which gives the best results.
The computed transverse velocities along with the corresponding sequence of magnetograms
have been used to calculate the helicity flux density.
The map of helicity flux density is shown in Fig. 12. In the region A the positive helicity dominates. However, in region B, both signs of helicity flux
are present.

To find how the helicity flux changed over time in each of these footpoints, we
have plotted the computed helicity flux separately for both the footpoints (A \& B) marked in boxes.
Figure~\ref{fig:13} shows the magnetic helicity flux for the box region A (top-left) and for the box region B (top-right).
The injected helicity flux is almost positive in sign in
box region A and it increased
to double the amount just after the filament eruption and remain unchanged even 8~h after
the filament eruption. But, in box region B the observed injected helicity flux
is predominantly positive during the first half on July 08, 2011 and turned negative after
16:00~UT on the same day. Though some fluctuations in its sign can still be seen.
This is also the time the negative flux
starts to emerge in the same region. The emerged flux would have injected opposite helicity flux
into the corona. The chirality of the filament located in this region is sinistral having
the positive helicity sign. It should be noted here that the injected helicity flux for
about 10~h duration before the filament eruption is opposite to the filament chirality.
The co-existence of both sign of helicity flux in the helicity flux density map support this
result (Fig.~\ref{fig:12}). After the filament eruption the dominance of the positive
helicity flux is restored in box region B.

The total accumulated helicity is positive in box region A and it was increasing
with time (Fig.~\ref{fig:13}~(bottom-left)). In box region B, though there was a net positive
helicity injection, the decrease in the helicity started about 10~h before the filament eruption
and increased afterward. This decrease in the net helicity is due to the injection of negative
helicity flux from the emerging flux region. This can be clearly seen in the magnetic
flux evolution plot in region B for the temporal coincidence. In either cases there is
a net positive helicity injection in both footpoints of the filament.

\section{Summary and discussions}
We studied a filament eruption that was formed in the vicinity of active region NOAA 11247.
While a large portion of filament was embedded in the quiet sun, the end footpoints were
rooted in the plage regions. The filament was located in the southern hemisphere with
`S-shaped' structure. The filament is of sinistral type and has positive helicity.
The filament eruption was initiated in the Western footpoint at around 23:20~UT,
with a projection speed of 56~km~s$^{-1}$ at the higher
chromosphere and transition region heights.

The Western end of the filament was rooted in negative polarity plage and the Eastern end
was located in the positive polarity plage region. There was a flux emergence in the nearby active
region. The increase in flux of the Western footpoint stopped a couple of hours before
the filament eruption. During the flux emergence, the negative flux pushed the
pre-existing positive flux of the active region further West. Soon after the
filament activation initiated there was an anti-clockwise rotational motion for about 6 min
in the Western footpoint of the filament. On the other hand the flux started to increase in
the Eastern footpoint of the filament a few hours before the filament eruption was initiated.
After the B4.7 class flare the increase in the flux stopped. Here also an anti-clockwise
rotational motion was observed at around 23:33~UT  -just after the filament
activation started in the Eastern footpoint of the filament. In Eastern and Western regions
of filament footpoint the computed magnetic helicity had positive sign. There was a rise
in the injected helicity flux a few minutes after the filament eruption in the Eastern
footpoint of the filament. But, in the Western footpoint both signs of helicity were
present, though it was completely positive at 10~h before the filament activation.
After the filament eruption, the dominance of positive magnetic helicity was restored.

The new magnetic flux emergence with opposite magnetic sign can destabilize the filament
by decreasing the magnetic tension of the overlying field \citep{Wang99}. Along with
the flux emergence the long duration converging motions observed in the vicinity of the
filament footpoints can reduce magnetic tension of the overlying field which can destabilize the
filament \citep{Zuccarello12a, Amari03, Amari10, Amari11}. The long lasting converging motion can also increase the axial
flux of the filament by transferring the overlying field into the underlying field of the
filament that eventually increase the outward magnetic pressure on the filament. The increase
in the magnetic pressure can push the filament to a height where the torus instability can
set in and drive the filament eruption as has been discussed in \cite{Zuccarello12b} and
\cite{Aulanier10, Demoulin10}.

On the other hand, the existence of the opposite
magnetic helicity in one of the footpoints of the filament can introduce a mutual interaction of
magnetic field having opposite magnetic helicity flux. In the Western footpoint of the
filament there was a dominant positive helicity flux injection in the beginning. Later, there
was a injection of opposite helicity flux into the existing positive helicity flux system.
\cite{Kusano03, Kusano04} concluded through numerical simulations that the coexistence of
opposite helicity flux would cause a reconnection while merging of different helicity flux
system takes place. The reconnection in the system can reduce the tension of the overlying field and thereby
pushing the filament to higher heights in the corona and thus driving the filament eruption as
discussed previously. Since we observed flux emergence, converging motion and the
opposite magnetic helicity flux in the filament footpoint, we believe one or more mechanisms
 may be playing a major role in initiating the filament eruption. Several of these possible mechanisms
are discussed in \cite{Romano11, Green11} and
\cite{Schmieder11}.

The more interesting event is the observed rotational motion in the footpoints of the
filament when the eruption had just set in. A possible explanation of the observed rotational
motion is advanced as follows. Once the filament eruption starts, there is an axial expansion of the filament flux rope.
If the footpoint of the flux rope is still anchored to the photosphere, the expansion of
the filament flux rope usually leads to the torque imbalance between the
photospheric footpoints and the coronal counter part of the expanded flux rope. The immediate
consequence of the torque imbalance is the generation of the shear flows at the photospheric
footpoints \citep{Parker74, Jockers78}. At the footpoint it is expected that unwinding
motion reduced the shear at the unexpanded portion and transferred a helicity into the
expanded portion.  This would increase the helicity flux of the same sign in the
expanded portion.
The filament with sinistral chirality has positive helicity flux in the system. As the filament
expands during the eruption, a change in the helicity flux from negative to positive soon
after the filament eruption in the Western footpoint suggests that there was a transfer of
positive helicity into the corona from the unexpanded part of the flux tube to the expanded part.
Later, the Eastern footpoint also showed the rotational motion
and there was a jump in the positive helicity after the initiation of the filament
eruption. This can be attributed to the expansion of the filament flux rope and
subsequent torque imbalance as has been suggested by \cite{Parker74, Chae03} and observed
by \cite{Smyrli10, Zuccarello11}.
This explains the observations of transient rotational motion in
the photospheric footpoints and increased positive magnetic helicity flux after the filament eruption. The 
rotational motion was no longer observed once the
footpoints were detached from the photosphere.

The filament eruption initiated just after the flux emergence stopped in one of the footpoints.
This kind of flux emergence followed by converging/shearing motions will
stress the overlaying fields and hence builds up energy in the corona \citep{Feynman95}.
In such condition,
a small perturbation can trigger the filament eruption.  
In dynamic plasma, the presence of small triggering agents is not uncommon.  
However, in each event the onset of instability could be different. What is required is to probe the magnetic shear
and stress in and around the filament region at least at two heights using vector magnetic
field measurements. In future, we plan to study magnetic field parameter changes using
vector magnetic field measurements taken at two different heights. It is also useful
to look for the rotational motion at the footpoints of the filaments and associated
changes in the magnetic helicity after the onset of filament eruption statistically. This can
provide an important clue to the generation of torque imbalance after the filament eruption
and help validate \cite{Parker74, Jockers78} results statistically.

\section{Acknowledgments}
We would like to thank all the referees for many valuable and insightful
comments which greatly helped us to improve the quality of the manuscripts.
Data and images are courtesy of NASA/SDO and the HMI
and AIA science teams. SDO/HMI is a joint effort of many teams and individuals
to whom we are greatly indebted for providing the data. We also acknowledge
BBSO consortia -a partner of global high resolution H$_{\alpha}$ network, for providing the data.

\end{document}